\documentclass[journal]{IEEEtran}
\usepackage[dvips]{graphicx}
\usepackage[cmex10]{amsmath}
\allowdisplaybreaks[4]
\usepackage{subfigure,cite,bbm,amssymb,multirow,amsthm,bm}
\usepackage{dsfont}
\usepackage[ruled, linesnumbered]{algorithm2e}
\usepackage{url,mathtools}
\usepackage{stfloats}
\usepackage{algpseudocode}

\makeatletter
\newlength \figwidth
\if@twocolumn
\else
\fi
\makeatother
\newcommand{\ind}{1\hspace{-2.2mm}1}

\usepackage{color}

\theoremstyle{plain}

\newtheorem{lem}{Lemma}
\newtheorem{prop}{Proposition}

\theoremstyle{definition}

\newtheorem{exmp}{Example}

\theoremstyle{remark}
\newtheorem{rem}{\bf Remark}

\newcommand{\av}{\ensuremath{\mathbf{y}}} 
\newcommand{\gx}{\ensuremath{\mathbf{g}}} 
\newcommand{\gp}{\ensuremath{\mathbf{v}}} 

\newcommand{\bfh}{\ensuremath{\mathbf{h}}}
\newcommand{\bfp}{\ensuremath{\mathbf{p}}}
\newcommand{\exv}{\ensuremath{\mathbb{E}}} 

\newcommand{\setp}{\ensuremath{\mathcal{P}}} 
\newcommand{\boxv}{\ensuremath{\mathcal{B}_{\mathbf{y}}}} 

\newcommand{\sg}{\ensuremath{\check{\bm{g}}}} 
\newcommand{\sph}{\ensuremath{\check{\bm{v}}}} 
\newcommand{\argmax}{\operatornamewithlimits{arg~max}}

\providecommand{\abs}[1]{\lvert#1\rvert}
\newtheoremstyle{amsassumstyle}
  {9pt}
  {9pt}
  {}
  {}
  {\bfseries}
  {.}
  {.5em}
  {\thmname{#1}\thmnumber{#2}\thmnote{ #3}}

\theoremstyle{amsassumstyle}
\newtheorem{assumption}{AS}

\begin{document}
\title{Cross-Layer Designs in Coded Wireless \\ Fading Networks with Multicast}
\author{Ketan Rajawat, Nikolaos Gatsis, and Georgios B. Giannakis,~\IEEEmembership{Fellow,~IEEE}%
\thanks{Manuscript received February 12, 2010; revised August 20, 2010. Work in this paper was supported by the NSF grants CCF-0830480,  ECCS-1002180,  and ECCS-0824007. Part of this paper has been presented at the 3rd IEEE Int. Workhop Wireless Network Coding, Boston, MA, June 2010.}
\thanks{ The authors are with the Department of Electrical and Computer Engineering, University of Minnesota, Minneapolis, MN 55455, USA. Tel/fax: (612)624-9510/625-2002, emails: \texttt{\{ketan,gatsisn,georgios\}@umn.edu}}}
\maketitle

\markboth{IEEE/ACM TRANSACTIONS ON NETWORKING (ACCEPTED)}{}

\begin{abstract}
A cross-layer design along with an optimal resource allocation framework is formulated for wireless fading networks, where the nodes are allowed to perform network coding. The aim is to jointly optimize end-to-end transport layer rates, network code design variables, broadcast link flows, link capacities, average power consumption, and short-term power allocation policies. As in the routing paradigm where nodes simply forward packets, the cross-layer optimization problem with network coding is non-convex in general. It is proved however, that with network coding, dual decomposition for multicast is optimal so long as the fading at each wireless link is a continuous random variable. This lends itself to provably convergent subgradient algorithms, which not only admit a layered-architecture interpretation but also optimally integrate network coding in the protocol stack. The dual algorithm is also paired with a scheme that yields near-optimal network design variables, namely multicast end-to-end rates, network code design quantities, flows over the broadcast links, link capacities, and average power consumption.  Finally, an asynchronous subgradient method is developed, whereby the dual updates at the physical layer can be affordably performed with a certain delay with respect to the resource allocation tasks in upper layers. This attractive feature is motivated by the complexity of the physical layer subproblem, and is an adaptation of the subgradient method suitable for network control.
\end{abstract}


\section{Introduction}
Traditional networks have always assumed nodes capable of only forwarding or replicating packets. For many types of networks however, this constraint is not inherently needed since the nodes can invariably perform encoding functions. Interestingly, even simple linear mixing operations can be powerful enough to enhance the network throughput, minimize delay, and decrease the overall power consumption \cite{yeung06},\cite{chou07}. For the special case of single-source multicast, which does not even admit a  polynomial-time solution within the routing framework \cite{multicast83}, linear network coding achieves the full network capacity \cite{yeung00}. In fact, the network flow description of multicast with random network coding adheres to only linear inequality constraints reminiscent of the corresponding description in unicast routing \cite{lunho06}. 

This encourages the use of network coding to extend several popular results in unicast routing framework to multicast without appreciable increase in complexity. Of particular interest is the resource allocation and cross-layer optimization task in wireless networks \cite{crosslay}, \cite{neely06}. The objective here is to maximize a network utility function subject to flow, rate, capacity and power constraints. This popular approach not only offers the flexibility of capturing diverse performance objectives, but also admits a layering interpretation, arising from different decompositions of the optimization problem \cite{mungchiang}.

This paper deals with cross-layer optimization of wireless \emph{multicast} networks that use network coding and operate over \emph{fading} links. The aim is to maximize a total network utility objective, and entails finding end-to-end rates, network code design variables, broadcast link flows, link capacities, average power consumption, and instantaneous power allocations.

Network utility maximization was first brought into coded networks in \cite{lunho06}, where the aim was to minimize a generic cost function subject only to flow and rate constraints. The optimal flow and rate variables may then be converted to a practical random network coding implementation using methods from \cite{lunmed04} and \cite{chouwu03}. Subsequent works extended this framework to include power, capacity, and scheduling constraints \cite{xiyeh-ton,ho08,traskov,amerimehr}. The interaction of network coding with the network and transport layers has also been explored in \cite{wu06,wuchiangkung,linli,hochiang07,hierch09}; in these works, networks with fixed link capacities are studied, and different decomposition techniques result in different types of layered architectures.

There are however caveats associated with the utility maximization problem in wireless networks. First, the power control and scheduling subproblems are usually non-convex. This implies that the dual decomposition of the overall problem, though insightful, is not necessarily optimal and does not directly result in a feasible primal solution. Second, for continuous fading channels, determining the power control policy is an infinite dimensional problem. Existing approaches in network coding consider either deterministic channels \cite{amerimehr,xiyeh-ton}, or, links with a finite number of fading states \cite{ho08,neely07,ho09}.

On the other hand, a recent result in unicast routing shows that albeit the non-convexity, the overall utility optimization problem has no duality gap for wireless networks with continuous fading channels~\cite{ale09}. As this is indeed the case in all real-life fading environments, the result promises the optimality of layer separation. In particular, it renders a dual subgradient descent algorithm for network design optimal~\cite{gatsis-journal}. 

The present paper begins with a formulation that jointly optimizes end-to-end rates, virtual flows, broadcast link flows, link capacities, average power consumption, and instantaneous power allocations in wireless fading multicast networks that use intra-session network coding (Section \ref{probfor}). The first contribution of this paper is to introduce a \emph{realistic physical layer model} formulation accounting for the capacity of broadcast links. The cross-layer problem is generally non-convex, yet it is shown to have zero duality gap (Section \ref{dualprop}). This result considerably broadens \cite{ale09} to \emph{coded multicast} networks with broadcast links. The zero duality gap is then leveraged in order to develop a subgradient descent algorithm that minimizes the dual function (Sections \ref{layersepar}, \ref{convoff}). The algorithm admits a natural layering interpretation, allowing optimal integration of network coding into the protocol stack.

In Section \ref{Sec:InNet}, the subgradient algorithm is modified so that the component of the subgradient that results from the physical layer power allocation may be delayed with respect to operations in other layers. This provably convergent asynchronous subgradient method and its \emph{online} implementation constitute the second major contribution. Unlike the algorithm in~\cite{gatsis-journal}, which is used for offline network optimization, the algorithm developed here is suitable for online network control. Convergence of asynchronous subgradient methods for dual minimization is known under diminishing stepsize~\cite{Kiwiel-Parallel}; the present paper proves results for constant stepsize. Near-optimal primal variables are also recovered by forming running averages of the primal iterates. This technique has also been used in synchronous subgradient methods for convex optimization; see e.g., \cite{nedic07} and references therein. Here, ergodic convergence results are established for the asynchronous scheme and the non-convex problem at hand. Finally, numerical results are presented in Section \ref{numerical}, and Section \ref{concl} concludes the paper.

\section{Problem Formulation}\label{probfor}
Consider a wireless network consisting of a set of terminals (nodes) denoted by $\mathcal{N}$. The broadcast property of the wireless interface is modeled by using the concept of hyperarcs. A {\it hyperarc} is a pair $(i,J)$ that represents a broadcast link from a node $i$ to a chosen set of nodes $J \subset \mathcal{N}$. The entire network can therefore be represented as a hypergraph $\mathcal{H} = (\mathcal{N},\mathcal{A})$, where $\mathcal{A}$ is the set of hyperarcs. The complexity of the model is determined by the choice of the set $\mathcal{A}$.
Let the neighbor-set $N(i)$ denote the set of nodes that node $i$ reaches. An exhaustive model might include all possible $2^{\abs{N(i)}}-1$ hyperarcs from node $i$. On the other hand, a simpler model might include only a smaller number of hyperarcs per node. A point-to-point model is also a special case when node $i$ has $\abs{N(i)}$ hyperarcs each containing just one receiver.

The present work considers a physical layer whereby the channels undergo random multipath fading. This model allows for opportunistically best schedules per channel realization. This is different from the link-level network models in \cite{ho08, ho09, lunho06, traskov}, where the hyperarcs are modeled as erasure channels. The next subsection discusses the physical layer model in detail.

\subsection{Physical Layer}\label{physec}
In the current setting, terminals are assumed to have a set of tones $\mathcal{F}$ available for transmission. Let $h^f_{ij}$ denote the power gain between nodes $i$ and $j$ over a tone $f \in \mathcal{F}$, assumed random, capturing fading effects. Let $\mathbf{h}$ represent the vector formed by stacking all the channel gains. The network operates in a time slotted fashion; the channel $\mathbf{h}$ remains constant for the duration of a slot, but is allowed to change from slot to slot. A slowly fading channel is assumed so that a large number of packets may be transmitted per time slot. The fading process is modeled to be stationary and ergodic.

Since the channel changes randomly per time slot, the optimization variables at the physical layer are the channel realization-specific power allocations $p_{iJ}^f(\mathbf{h})$ for all hyperarcs $(i,J) \in \mathcal{A}$, and tones $f \in \mathcal{F}$. For convenience, these power allocations are stacked in a vector $\mathbf{p}(\mathbf{h})$. {Instantaneous power allocations may adhere to several scheduling and mask constraints, and these will be generically denoted by a bounded set $\Pi$ such that $\mathbf{p}(\mathbf{h}) \in \Pi$}.
The long-term average power consumption by a node $i$  is given by
\begin{equation}\label{pow}
p_i = \mathbb{E}\left[\sum_{f}\sum_{J : (i,J)\in \mathcal{A}}p^f_{iJ}(\mathbf{h})\right]
\end{equation}
where $\mathbb{E}[.]$ denotes expectation over the stationary channel distribution. 

For slow fading channels, the information-theoretic capacity of a hyperarc $(i,J)$ is defined as the maximum rate at which \emph{all} nodes in $J$ receive data from $i$ with vanishing probability of error in a given time slot. This capacity depends on the instantaneous power allocations $\mathbf{p}(\mathbf{h})$ and channels $\mathbf{h}$. A generic bounded function $C_{iJ}^f(\mathbf{p}(\mathbf{h}),\mathbf{h})$ will be used to describe this mapping. Next we give two examples of the functional forms of $C_{iJ}^f(\cdot)$ and $\Pi$.
\begin{exmp}
Conflict graph model: The power allocations $p_{iJ}^f$ adhere to the spectral mask constraints
\begin{equation}
0 \leq p_{iJ}^f \leq p_{\max}^f.
\end{equation}
However, only conflict-free hyperarc are allowed to be scheduled for a given $\bfh$. Specifically, power may be allocated to hyperarcs $(i_1,J_1)$ and $(i_2,J_2)$ if and only if \cite{traskov}
\begin{enumerate}
	\item[i)] $i_1\neq i_2$;
	\item[ii)] $i_1 \notin J_2$ and $i_2 \notin J_1$ (half-duplex operation); and
	\item[iii-a)] $J_1 \cap J_2 = \emptyset$ (primary interference), or additionally,
	\item[iii-b)] $J_1 \cap N(i_2) = J_2 \cap N(i_1) = \emptyset$ (secondary interference).		
\end{enumerate}
The set $\Pi$ therefore consists of all possible power allocations that satisfy the previous properties.

Due to hyperarc scheduling, all transmissions in the network are interference free. The signal-to-noise ratio (SNR) at a node $j\in J$ is given by
\begin{equation}\label{snr}
\Gamma_{iJj}^f(\mathbf{p}(\mathbf{h}),\mathbf{h}) = p_{iJ}^f(\mathbf{h})h^f_{ij}/N_j
\end{equation}
where $N_j$ is the noise power at $j$. In a broadcast setting, the maximum rate of information transfer from $i$ to \emph{each} node in $J$ is
\begin{equation}\label{harccap}
C_{iJ}^f(\mathbf{p}(\mathbf{h}),\mathbf{h}) = \min_{j\in J} \log(1+\Gamma_{iJj}^f(\mathbf{p}(\mathbf{h}),\mathbf{h})).
\end{equation}
A similar expression can be written for the special case of point-to-point links by substituting hyperarcs $(i,J)$ by arcs $(i,j)$ in the expression for $\Gamma_{iJj}^f(\mathbf{p}(\mathbf{h}),\mathbf{h})$. 

For slow-fading channels, Gaussian codebooks with sufficiently large block lengths achieve this capacity in every time slot. More realistically, an SNR penalty term $\rho$ can be included to account for finite-length practical codes and adaptive modulation schemes, so that
\begin{equation}\label{amccap}
C_{iJ}^f(\mathbf{p}(\mathbf{h}),\mathbf{h}) = \min_{j\in J} \log\left(1+\Gamma_{iJj}^f(\mathbf{p}(\mathbf{h}),\mathbf{h})/\rho\right). 
\end{equation} 
The penalty term is in general a function of the target bit error rate.
\end{exmp}
\begin{exmp}
Signal-to-interference-plus-noise-ratio (SINR) model: Here, the constraint set $\Pi$ is simply a box set $\mathcal{B}_\mathbf{p}$,
\begin{equation}
\Pi = \mathcal{B}_\bfp := \{p_{iJ}^f | 0 \leq p_{iJ}^f \leq p_{\max}^f~ \forall~ \text{$(i,J) \in \mathcal{A}$ and $f\in \mathcal{F}$ } \}.
\end{equation}
The set $\mathcal{B}_\bfp$ could also include (instantaneous) sum-power constraints per node. The capacity is expressed as in \eqref{harccap} or \eqref{amccap}, but now the SNR is replaced by the SINR, given by
\begin{equation}
\Gamma_{iJj}^f(\mathbf{p}(\mathbf{h}),\mathbf{h}) = \left. p_{iJ}^f(\mathbf{h})h_{ij}^f \right/ \left(N_j +  I_{ij,f}^{\mathrm{int}}+  I_{j,f}^{\mathrm{self}}+  I_{iJj,f}^{\mathrm{broad}}\right).
\end{equation}
The denominator consists of the following terms:
\begin{subequations}
\begin{itemize}
	\item Interference from other nodes' transmissions to node $j$
	\begin{equation}
	I_{ij,f}^{\mathrm{int}} = \sum_{\substack{ (k,M) \in \mathcal{A} : j\in M, \\k\neq j, k\neq i}} p_{kM}^f(\mathbf{h})h_{kj}^f.
	\end{equation}
	\item ``Self-interference'' due to transmissions of node $j$
	\begin{equation}
	I_{j,f}^{\mathrm{self}} = h_{jj} \sum_{M : (j,M) \in \mathcal{A}}p_{jM}^f(\mathbf{h}).
	\end{equation}
	This term is introduced to encourage half-duplex operation by setting $h_{jj}$ to a large value.
	\item ``Broadcast-interference'' from transmissions of node $i$ to other hyperarcs
	\begin{equation}
	I_{iJj,f}^{\mathrm{broad}} = \beta h_{ij}^f \sum_{\substack{M : (i,M) \in \mathcal{A}\\M\neq J}}p_{iM}^f(\mathbf{h}).
	\end{equation}
	This term is introduced to force node $i$ to transmit at most over a single hyperarc, by setting $\beta$ to a large value.
\end{itemize}
\end{subequations}
The previous definitions ignore interference from non-neighboring nodes. However, they can be readily extended to include more general interference models.
\end{exmp}

The link layer capacity is defined as the long-term average of the total instantaneous capacity, namely,
\begin{equation}\label{cap}
c_{iJ}:= \mathbb{E}\left[\sum_fC_{iJ}^f(\mathbf{p}(\mathbf{h}),\mathbf{h})\right].
\end{equation}
This is also called ergodic capacity and represents the maximum average data rate available to the link layer.

\subsection{Link Layer and Above}
The network supports multiple multicast sessions indexed by $m$, namely $\mathcal{S}_m := (s^m, T^m, a^m)$, each associated with a source node $s^m$, sink nodes $T^m \subset \mathcal{N}$, and an average flow rate $a^m$ from $s^m$ to each $t \in T^m$. The value $a^m$ is the average rate at which the network layer of source terminal $s^m$ admits packets from the transport layer. Traffic is considered elastic, so that the packets do not have any short-term delay constraints. 

Network coding is a generalization of routing since the nodes are allowed to code packets together rather than simply forward them. This paper considers intra-session network coding, where only the traffic belonging to the same multicast session is allowed to mix. Although better than routing in general, this approach is still suboptimal in terms of achieving the network capacity. However, general (inter-session) network coding is difficult to characterize or implement since neither the capacity region nor efficient network code designs are known \cite[Part II]{yeung06}. On the other hand, a simple linear coding strategy achieves the full capacity region of intra-session network coding \cite{yeung00}. 

The network layer consists of endogenous flows of coded packets over hyperarcs. Recall that the maximum average rate of transmission over a single hyperarc cannot exceed $c_{iJ}$. Let the coded packet-rate of a multicast session $m$ over hyperarc $(i,J)$ be $z_{iJ}^m$ (also referred to as the subgraph or broadcast link flow). The link capacity constraints thus translate to
\begin{equation}\label{arccap}
\sum_{m}z_{iJ}^m \leq c_{iJ} \quad \forall (i,J) \in \mathcal{A}.
\end{equation}

To describe the intra-session network coding capacity region, it is commonplace to use the concept of \emph{virtual flow} between terminals $i$ and $j$ corresponding to each session $m$ and sink $t \in T^{m}$ with average rate $x_{ij}^{mt}$. These virtual flows are defined only for neighboring pairs of nodes i.e., $(i,j)\in \mathcal{G}:=\{(i,j)|(i,J) \in \mathcal{A}, j\in J\}$. The virtual flows satisfy the flow-conservation constraints, namely,
\begin{equation}\label{flowcon}
\sum_{j : (i,j)\in\mathcal{G}} x_{ij}^{mt} - \sum_{j : (j,i)\in\mathcal{G}}x_{ji}^{mt} = \sigma_i^{m} := \begin{cases} a^m & \text{if }i = s^m, \\
-a^m & \text{if }i = t, \\
0 & \text{otherwise}\end{cases}
\end{equation}
for all $m$, $t \in T^m$, and $i \in \mathcal{N}$. Hereafter, the set of equations for $i = t$ will be omitted because they are implied by the remaining equations.

The broadcast flows $z_{iJ}^m$ and the virtual flows $x_{ij}^{mt}$ can be related using results from the lossy-hyperarc model of \cite{lunho06, traskov}. Specifically, \cite[eq. (9)]{traskov} relates the virtual flows and subgraphs, using the fraction $b_{iJK} \in [0,1]$ of packets injected into the hyperarc $(i,J)$ that reach the set of nodes $K \subset N(i)$. Recall from Section \ref{physec}, that here the instantaneous capacity function $C_{iJ}^f(\cdot)$ is defined such that all packets injected into the hyperarc $(i,J)$ are received by every node in $J$. Thus in our case, $b_{iJK} = 1$ whenever $K \cap J \neq \emptyset$ and consequently,
\begin{equation}\label{maxflow}
\sum_{j\in K} x^{mt}_{ij} \leq \sum_{\substack{J : (i,J)\in \mathcal{A} \\
J\cap K \neq \emptyset}}  z_{iJ}^m\quad\forall K \subset N(i),i\in \mathcal{N},m,t\in T^m.
\end{equation}

Note the difference with \cite{traskov} where at every time slot, packets are injected into a fixed set of hyperarcs at the same rate. The problem in \cite{traskov} is therefore to find a schedule of hyperarcs that do not interfere (the non-conflicting hyperarcs). The same schedule is used at every time slot; however, only a random subset of nodes receive the injected packets in a given slot. Instead here, the hyperarc selection is part of the power allocation problem at the physical layer, and is done for every time slot. The transmission rate (or equivalently, the channel coding redundancy) is however appropriately adjusted so that all the nodes in the selected hyperarc receive the data. 

In general, for any feasible solution to the set of equations \eqref{arccap}-\eqref{maxflow}, a network code exists that supports the corresponding exogenous rates $a^m$ \cite{lunho06}. This is because for each multicast session $m$, the maximum flow between $s^m$ and $t  \in T^m$ is $a^m$, and is therefore achievable \cite[Th. 1]{yeung00}. Given a feasible solution, various network coding schemes can be used to achieve the exogenous rates. Random network coding based implementations such as those proposed in \cite{lunmed04} and \cite{chouwu03}, are particularly attractive since they are fully distributed and require little overhead. These schemes also handle any residual errors or erasures that remain due to the physical layer.

The system model also allows for a set of ``box constraints'' that limit the long-term powers, transport layer rates, broadcast link flow rates, virtual flow rates as well as the maximum link capacities. Combined with the set $\Pi$, these constraints can be compactly expressed as
\begin{align}
\mathcal{B} := \{\mathbf{y},\mathbf{p}(\mathbf{h}) |~ & \bfp(\mathbf{h}) \in \Pi , ~~ 0\leq p_i \leq p^{\max}_{i}, \nonumber\\
& a^m_{\min} \leq a^m \leq a^m_{\max} , ~~ 0\leq c_{iJ}\leq c^{\max}_{iJ}, \nonumber\\
& 0\leq z^{m}_{iJ} \leq z^{\max}_{iJ}, ~~ 0\leq x_{ij}^{mt} \leq x^{\max}_{ij}\}. 
\end{align}
Here $\mathbf{y}$ is a super-vector formed by stacking all the average rate and power variables, that is, $a^m$, $z_{iJ}^m$, $x_{ij}^{mt}$, $c_{iJ}$, and $p_i$. Parameters with min/max subscripts or superscripts denote prescribed lower/upper bounds on the corresponding variables.

\subsection{Optimal Resource Allocation}
A common objective of the network optimization problem is maximization of the exogenous rates $a^m$ and minimization of the power consumption $p_i$. Towards this end, consider increasing and concave utility functions $U_m(a^m)$ and convex cost functions $V_i(p_i)$ so that the overall objective function $f(\mathbf{y}) = \sum_m U_m(a^m) - \sum_iV(p_i)$ is concave. 
For example, the utility function can be the logarithm of session rates and the cost function can be the squared average power consumption. The network utility maximization problem can be written as
\begin{subequations}\label{netopt}
\begin{align}
\mathsf{P} =& \max_{(\mathbf{y},\mathbf{p}(\mathbf{h}))\in\mathcal{B}} \sum_m U_m(a^m) - \sum_{i}V_i(p_i) \\
\text{s. t.}~&\sigma_i^{m} \leq \sum_{(i,j)\in\mathcal{G}} x_{ij}^{mt} - \sum_{(j,i)\in\mathcal{G}}x_{ji}^{mt} && \hspace{-0.5cm} \forall ~ m, i \neq t, t\in T^m \label{flowconr}\\
&\hspace{-.3cm} \sum_{j\in K} x^{mt}_{ij} \leq \sum_{\substack{J : (i,J)\in \mathcal{A} \\
J\cap K \neq \emptyset}} z_{iJ}^m && \hspace{-1.5cm} \forall~ K \subset N(i), m, t\in T^m \label{maxflowr}\\
&\hspace{-.3cm}\sum_{m} z_{iJ}^m \leq c_{iJ} && \hspace{-0.7cm} \forall~ (i,J)\in\mathcal{A}  \label{arccapr}\\
&\hspace{-.3cm}c_{iJ} \leq \mathbb{E}\left[\sum_fC_{iJ}^f(\mathbf{p}(\mathbf{h}),\mathbf{h})\right] && \hspace{-0.7cm} \forall~ (i,J)\in\mathcal{A} \label{capr}\\
&\hspace{-.3cm}\mathbb{E}\left[\sum_{f}\sum_{J : (i,J)\in \mathcal{A}}p^f_{iJ}(\mathbf{h})\right] \leq p_i  && \hspace{-0.7cm}\label{powr}
\end{align}
\end{subequations}
where $i\in\mathcal{N}$. Note that constraints \eqref{pow}, \eqref{cap} and \eqref{flowcon} have been relaxed without increasing the objective function.  For instance, the relaxation of \eqref{flowcon} is equivalent to allowing each node to send at a higher rate than received, which amounts to adding virtual sources at all nodes $i \neq t$. However, adding virtual sources does not result in an increase in the objective function because the utilities $U_m$ depend only on the multicast rate $a^m$.

The solution of the optimization problem \eqref{netopt} gives the throughput $a^m$ that is achievable using optimal virtual flow rates $x_{ij}^{mt}$ and power allocation policies $\mathbf{p}(\mathbf{h})$. These virtual flow rates are used for network code design. When implementing coded networks in practice, the traffic is generated in packets and stored at nodes in queues (and virtual queues for virtual flows) \cite{chouwu03}. The constraints in \eqref{netopt} guarantee that all queues are stable. 

Optimization problem \eqref{netopt} is non-convex in general, and thus difficult to solve. For example, in the conflict graph model, the constraint set $\Pi$ is discrete and non-convex, while in the SINR-model, the capacity function $C_{iJ}^f(\mathbf{p}(\mathbf{h}),\mathbf{h})$ is a non-concave function of $\mathbf{p}(\mathbf{h})$; see e.g., \cite{Luo-Zhang-DSM},\cite{crosslay}. The next section analyzes the Lagrangian dual of \eqref{netopt}. 

\section{Optimality of Layering}\label{subgrad}
This section shows that~\eqref{netopt} has zero duality gap, and solves the dual problem via subgradient descent iterations. The purpose here is two-fold: $i$) to describe a layered architecture in which linear network coding is optimally integrated; and $ii$) to set the basis for a network implementation of the subgradient method, which will be developed in Section~\ref{Sec:InNet}.

\subsection{Duality Properties}\label{dualprop}
Associate Lagrange multipliers $\nu_{i}^{mt}$, $\eta_{iK}^{mt}$, $\xi_{iJ}$, $\lambda_{iJ}$ and $\mu_i$ with the flow constraints \eqref{flowconr}, the union of flow constraints \eqref{maxflowr}, the link rate constraints \eqref{arccapr}, the capacity constraints \eqref{capr}, and the power constraints \eqref{powr}, respectively. Also, let $\boldsymbol{\zeta}$ be the vector formed by stacking these Lagrange multipliers in the aforementioned order. Similarly, if inequalities \eqref{flowconr}--\eqref{powr} are rewritten with zeros on the right-hand side, the vector  $\mathbf{q}(\mathbf{y},\mathbf{p}(\mathbf{h}))$ collects all the terms on the left-hand side of the constraints. The Lagrangian can therefore be written as
\begin{align}
{\mathcal{L}}(\mathbf{y},\mathbf{p}(\mathbf{h}),\boldsymbol{\zeta}) &= \sum_m U_m(a^m) - \sum_{i\in\mathcal{N}}V_i(p_i) - \boldsymbol{\zeta}^T\mathbf{q}(\mathbf{y},\mathbf{p}(\mathbf{h})). \label{lagrangian}
\end{align}
The dual function and the dual problem are, respectively,
\begin{align}
\varrho(\boldsymbol{\zeta}) &:= \max_{(\mathbf{y},\mathbf{p}(\mathbf{h}))\in\mathcal{B}} {\mathcal{L}}(\mathbf{y},\mathbf{p}(\mathbf{h}),\boldsymbol{\zeta}) \label{dualf}\\
\mathsf{D} &= \min_{\boldsymbol{\zeta}\geq \mathbf{0}} \varrho(\boldsymbol{\zeta}). \label{dual}
\end{align}
Since \eqref{capr} may be a non-convex constraint, the duality gap is in general, non-zero; i.e., $\mathsf{D} \geq \mathsf{P}$. Thus, solving \eqref{dual} yields an upper bound on the optimal value $\mathsf{P}$ of \eqref{netopt}. In the present formulation however, we have the following interesting result.

\begin{prop}\label{zdualp}
If the fading is continuous, then the duality gap is exactly zero, i.e., 
$\mathsf{P} = \mathsf{D}$.
\end{prop}

A generalized version of Proposition \ref{zdualp}, including a formal definition of continuous fading, is provided in Appendix \ref{zerodual} and connections to relevant results are made. The essential reason behind this strong duality is that the set of ergodic capacities resulting from all feasible power allocations is convex.

The requirement of continuous fading channels is not limiting since it holds for all practical fading models, such as Rayleigh, Rice, or Nakagami-$m$. Recall though that the dual problem is always convex. The subgradient method has traditionally been used to approximately solve \eqref{dual}, and also provide an intuitive layering interpretation of the network optimization problem \cite{mungchiang}. The zero duality gap result is remarkable in the sense that it renders this layering optimal. 

A corresponding result for unicast routing in uncoded networks has been proved in \cite{ale09}. The fact that it holds for coded networks with broadcast links, allows optimal integration of the network coding operations in the wireless protocol stack. The next subsection deals with this subject.

\subsection{Subgradient Algorithm and Layer Separability}\label{layersepar}
The dual problem \eqref{dual} can in general be solved using the subgradient iterations \cite[Sec.~8.2]{Bertsekas-ConvexAnalysis} indexed by $\ell$
\begin{subequations}\label{updprdu}
\begin{align}
(\mathbf{y}(\ell),\mathbf{p}(\mathbf{h};\ell)) &\in \argmax_{(\mathbf{y},\mathbf{p}(\mathbf{h}))\in\mathcal{B}} {\mathcal{L}}(\mathbf{y},\mathbf{p}(\mathbf{h}),\boldsymbol{\zeta}(\ell)) \label{updprim}\\
\boldsymbol{\zeta}(\ell+1) &= \left[\boldsymbol{\zeta}(\ell) + \epsilon\mathbf{q}(\mathbf{y}(\ell),\mathbf{p}(\mathbf{h};\ell))\right]^{+}\label{updsub}
\end{align}
\end{subequations}
where $\epsilon$ is a positive constant stepsize, and $[.]^{+}$ denotes projection onto the nonnegative orthant. The inclusion symbol ($\in$) allows for potentially multiple maxima. In \eqref{updsub}, $\mathbf{q}(\mathbf{y}(\ell),\mathbf{p}(\mathbf{h};\ell))$ is a subgradient of the dual function $\varrho(\boldsymbol{\zeta})$ in \eqref{dualf} at $\boldsymbol{\zeta}(\ell)$. Next, we discuss the operations in \eqref{updprdu} in detail.

For the Lagrangian obtained from \eqref{lagrangian}, the maximization in \eqref{updprim} can be separated into the following subproblems
\begin{subequations}\label{updprime}
\begin{align}
\hspace{-.2cm} a_i^m(\ell) &\in \argmax_{a^m_{\min}\leq a^m \leq a^m_{\max}} \left[U_m(a^m) - \sum_{t\in T^m}{\nu_{s_m}^{mt}(\ell)}a^m\right] \label{rate}\\
\hspace{-.2cm}z_{iJ}^m(\ell) &\in \argmax_{0\leq z^m_{iJ} \leq z_{iJ}^{\max}} \left[\sum_{\substack{K \subset N(i) \\
K\cap J \neq \emptyset}}\sum_{t\in T^m} {\eta_{iK}^{mt}(\ell)} - \xi_{iJ}(\ell)\right]z_{iJ}^m \label{acflow}\\
\hspace{-.2cm}x_{ij}^{mt}(\ell) &\in \argmax_{ 0\leq x_{ij}^{mt} \leq x^{\max}_{ij}} \nonumber\\ 
&\hspace{-.5cm}\left[{\nu_i^{mt}}(\ell)\ind_{i\neq t}-{\nu_j^{mt}}(\ell)\ind_{j\neq t}-\sum_{\substack{K\subset N(i)\\ j\in K}} {\eta_{iK}^{mt}}(\ell)\right]x_{ij}^{mt} \label{vflow}\\
\hspace{-.2cm}c_{iJ}(\ell) &\in \argmax_{0\leq c_{iJ}\leq c_{iJ}^{\max}} \left[\xi_{iJ}(\ell)-\lambda_{iJ}(\ell)\right] c_{iJ} \label{capacity}\\
\hspace{-.2cm}p_i(\ell) &\in \argmax_{0\leq p_i \leq p_{i}^{\max}}\left[ \mu_i(\ell) p_i - V_i(p_i) \right] \label{power}\\
\hspace{-.2cm}\mathbf{p}(\mathbf{h};\ell) &\in \argmax_{ \bfp(\mathbf{h}) \in \Pi} \sum_{f,(i,J)\in\mathcal{A}} \gamma_{iJ}^f(\mathbf{p}(\mathbf{h}),\mathbf{h},\bm\zeta)\label{powcon}
\shortintertext{where}
\gamma_{iJ}^f&(\mathbf{p}(\mathbf{h}),\mathbf{h},\bm\zeta) := \lambda_{iJ}C_{iJ}^f(\mathbf{p}(\mathbf{h}),\mathbf{h}) - \mu_i p_{iJ}^f(\mathbf{h})\label{gamma} 
\end{align}
\end{subequations}
and $\ind_X$ is the indicator function, which equals one if the expression $X$ is true, and zero otherwise.

The physical layer subproblem \eqref{powcon} implies per-fading state separability. Specifically, instead of optimizing over the class of power control policies, \eqref{powcon} allows solving for the optimal power allocation for each fading state; that is,
\begin{align}
\mathsf{P}(\mathbf{p}(\mathbf{h})) &= \max_{\bfp(\mathbf{h}) \in \Pi} \mathbb{E}\left[\sum_{f,(i,J)\in\mathcal{A}}\gamma_{iJ}^f(\mathbf{p}(\mathbf{h}),\mathbf{h},\bm\zeta) \right] \nonumber\\
&\hspace{-1.5cm}= \mathbb{E} \left[\max_{\bfp(\mathbf{h}) \in \Pi} \sum_{f,(i,J)\in\mathcal{A}} \gamma_{iJ}^f(\mathbf{p}(\mathbf{h}),\mathbf{h},\bm\zeta) \right]. \label{powprob}
\end{align}

Note that problems \eqref{rate}--\eqref{power} are convex and admit efficient solutions. The per-fading state power allocation subproblem \eqref{powcon} however, may not necessarily be convex. For example, under the conflict graph model (cf.\ Example 1), the number of feasible power allocations may be exponential in the number of nodes. Finding an allocation that maximizes the objective function in \eqref{powprob} is equivalent to the NP-hard maximum weighted hyperarc matching problem \cite{traskov}. Similarly, the capacity function and hence the objective function for the SINR model (cf.\ Example 2) is non-convex in general, and may be difficult to optimize.

This separable structure allows a useful layered interpretation of the problem. In particular, the transport layer sub-problem \eqref{rate} gives the optimal exogenous rates allowed into the network; the network flow sub-problem \eqref{acflow} yields the endogenous flow rates of coded packets on the hyperarcs; and the virtual flow sub-problem \eqref{vflow} is responsible for determining the virtual flow rates between nodes and therefore the network code design. Likewise, the capacity sub-problem \eqref{capacity} yields the link capacities and the power sub-problem \eqref{power} provides the power control at the data link layer.

The layered architecture described so far also allows for optimal integration of network coding into the protocol stack. Specifically, the broadcast and virtual flows optimized respectively in \eqref{acflow} and \eqref{vflow}, allow performing the combined routing-plus-network coding task at network layer. An implementation such as the one in \cite{chouwu03} typically requires queues for both broadcast as well as virtual flows to be maintained here.

Next, the subgradient updates of \eqref{updsub} become
\begin{subequations}\label{updsube}
\begin{align}
\nu_i^{mt}(\ell+1) &= \left[\nu_i^{mt}(\ell) + \epsilon\check{q}_{\nu}^{imt}(\ell)\right]^{+}\label{nut}\\
\eta_{iK}^{mt}(\ell+1) &= \left[\eta_{iK}^{mt}(\ell) + \epsilon\check{q}_{\eta}^{iKmt}(\ell)\right]^{+}\label{etat}\\
\xi_{iJ}(\ell+1) &= \left[\xi_{iJ}(\ell) + \epsilon\check{q}_{\xi}^{iJ}(\ell)\right]^{+}\label{xit}\\
\lambda_{iJ}(\ell+1) &= \left[\lambda_{iJ}(\ell) + \epsilon\check{q}_{\lambda}^{iJ}(\ell)\right]^{+}\label{lambdat}\\
\mu_{i}(\ell+1) &= \left[\mu_{i}(\ell) + \epsilon\check{q}_{\mu}^{i}(\ell)\right]^{+}\label{mut}
\end{align}
\end{subequations}
where $\check{q}(\ell)$ are the subgradients at index $\ell$ given by
\begin{subequations}\label{subgrads}
\begin{align}
\check{q}_{\nu}^{imt}(\ell) &= \sigma_i^m(\ell) + \sum_{(i,j)\in\mathcal{G}}x_{ji}^{mt}(\ell) - \sum_{(j,i)\in\mathcal{G}} x_{ij}^{mt}(\ell)\\
\check{q}_{\eta}^{iKmt}(\ell)&= \sum_{j \in K} x^{mt}_{ij}(\ell) - \sum_{\substack{J : (i,J)\in\mathcal{A}\\ J \cap K \neq \emptyset}}z^m_{iJ}(\ell)\\
\check{q}_{\xi}^{iJ}(\ell)&=\sum_{m} z_{iJ}^m(\ell) - c_{iJ}(\ell)\\
\check{q}_{\lambda}^{iJ}(\ell)&=c_{iJ}(\ell) - \mathbb{E}\left[\sum_fC_{iJ}^f(\mathbf{p}(\mathbf{h};\ell),\mathbf{h})\right]\label{capexp}\\
\check{q}_{\mu}^{i}(\ell)&=\mathbb{E}\left[\sum_{f}\sum_{J : (i,J)\in \mathcal{A}}p^f_{iJ}(\mathbf{h};\ell)\right] - p_i(\ell).\label{powexp}
\end{align}
\end{subequations}
The physical layer updates \eqref{lambdat} and \eqref{mut} are again complicated since they involve the $\mathbb{E}[.]$ operations of \eqref{capexp} and \eqref{powexp}. These expectations can be acquired via Monte Carlo simulations by solving \eqref{powcon} for realizations of $\mathbf{h}$ and averaging over them. These realizations can be independently drawn from the distribution of $\bfh$, or they can be actual channel measurements. In fact, the latter is implemented in Section \ref{Sec:InNet} on the fly during network operation. 

\subsection{Convergence Results}\label{convoff}
This subsection provides convergence results for the subgradient iterations \eqref{updprdu}. Since the primal variables $(\mathbf{y}, \mathbf{p}(\mathbf{h}))$ and the capacity function $C_{iJ}^f(.)$ are bounded, it is possible to define an upper bound $G$ on the subgradient norm; i.e., $\left\|\mathbf{q}(\mathbf{y}(\ell),\mathbf{p}(\mathbf{h};\ell))\right\| \leq G$ for all $\ell \geq 1$. 
\begin{prop}\label{dualconv}
For the subgradient iterations in \eqref{updprime} and \eqref{updsube}, the best dual value converges to $\mathsf{D}$ upto a constant; i.e.,
\begin{equation}
\lim_{s\rightarrow \infty}\min_{1\leq \ell \leq s} \varrho(\boldsymbol{\zeta}(\ell)) \leq \mathsf{D} + \frac{\epsilon G^2}{2}.
\end{equation}
\end{prop}

This result is well known for dual (hence, convex) problems \cite[Prop.~8.2.3]{Bertsekas-ConvexAnalysis}. However, the presence of an infinite-dimensional variable $\mathbf{p}(\mathbf{h})$ is a subtlety here. A similar case is dealt with in \cite{ale09} and Proposition \ref{dualconv} follows from the results there.

Note that in the subgradient method \eqref{updprdu}, the sequence of primal iterates $\{\mathbf{y}(\ell)\}$ does not necessarily converge. However, a primal running average scheme can be used for finding the optimal primal variables $\mathbf{y}^{*}$ as summarized next. Recall that $f(\mathbf{y})$ denotes the objective function $\sum_m U_m(a^m) - \sum_i V_i(p_i)$.
\begin{prop} \label{primconv} For the running average of primal iterates  
\begin{equation}\label{runav}
\bar{\mathbf{y}}(s) := \frac{1}{s}\sum_{\ell=1}^{s}\mathbf{y}(\ell).
\end{equation}
the following results hold:
\begin{enumerate}
	\item[a)] There exists a sequence $\{\mathring{\mathbf{p}}(\mathbf{h};s)\}$ such that $(\bar{\mathbf{y}}(s),\mathring{\mathbf{p}}(\mathbf{h};s)) \in \mathcal{B}$, and also
\begin{equation}\label{syfeas}
\lim_{s\rightarrow\infty}\left\|\left[\mathbf{q}(\bar{\mathbf{y}}(s),\mathring{\mathbf{p}}(\mathbf{h};s))\right]^{+}\right\| = 0.
\end{equation}
	\item[b)] The sequence $f(\bar{\mathbf{y}}(s))$ converges in the sense that
	\vspace{-0.3cm}
\begin{subequations}\label{primconst}
\begin{align}
\liminf_{s\rightarrow\infty} f(\bar{\mathbf{y}}(s)) &\geq \mathsf{P}-\frac{\epsilon G^2}{2}\\
\text{and~~}\limsup_{s\rightarrow\infty} f(\bar{\mathbf{y}}(s)) &\leq \mathsf{P}.
\end{align}
\end{subequations}
\end{enumerate}
\end{prop}

Equation \eqref{syfeas} asserts that the sequence $\{\bar{\mathbf{y}}(\ell)\}$ together with an associated $\{\mathring\bfp(\mathbf{h};\ell)\}$ becomes asymptotically feasible. Moreover, \eqref{primconst} explicates the asymptotic suboptimality as a function of the stepsize, and the bound on the subgradient norm. Proposition \ref{primconv} however, does not provide a way to actually find $\{\mathring\bfp(\mathbf{h};\ell)\}$.

Averaging of the primal iterates is a well-appreciated method to obtain optimal primal solutions from dual subgradient methods in convex optimization \cite{nedic07}. Note though that the primal problem at hand is non-convex in general. Results related to Proposition~\ref{primconv} are shown in~\cite{gatsis-journal}.
Proposition~\ref{primconv} follows in this paper as a special case result for a more general algorithm allowing for asynchronous subgradients and suitable for online network control, elaborated next.


\section{Subgradient Algorithm for Network Control}\label{Sec:InNet}
The algorithm in Section \ref{layersepar} finds the optimal operating point of \eqref{netopt} in an offline fashion. In the present section, the subgradient method is adapted so that it can be used for resource allocation during network operation.

The algorithm is motivated by Proposition \ref{primconv} as follows. The exogenous arrival rates $a^m(\ell)$ generated by the subgradient method [cf.\ \eqref{rate}] can be used as the instantaneous rate of the traffic admitted at the transport layer at time $\ell$. Then, Proposition \ref{primconv} guarantees that the long-term average transport layer rates will be optimal. Similar observations can be made for other rates in the network. 

More generally, an online algorithm with the following characteristics is desirable.
\begin{itemize}
	\item Time is divided in slots and each subgradient iteration takes one time slot. The channel is assumed to remain invariant per slot but is allowed to vary across slots.
	\item Each layer maintains its set of dual variables, which are updated according to \eqref{updsube} with a constant stepsize $\epsilon$. 
	\item The instantaneous transmission and reception rates at the various layers are set equal to the primal iterates at that time slot, found using \eqref{updprime}.
	\item Proposition \ref{primconv} ensures that the long-term average rates are optimal.
\end{itemize}

For network resource allocation problems such as those described in \cite{lunho06}, the subgradient method naturally lends itself to an online algorithm with the aforementioned properties. This approach however cannot be directly extended to the present case because the dual updates \eqref{lambdat}--\eqref{mut} require an expectation operation, which needs prior knowledge of the exact channel distribution function for generation of independent realizations of $\mathbf{h}$ per time slot. Furthermore, although Proposition \ref{primconv} guarantees the existence of a sequence of feasible power variables $\mathring{\mathbf{p}}(\mathbf{h};s)$, 
it is not clear if one could find them since the corresponding running averages do not necessarily converge. 

Towards adapting the subgradient method for network control, recall that the subgradients $\check{q}_\lambda^{iJ}$ and $\check{q}_\mu^i$ involve the following summands that require the expectation operations [cf.\ \eqref{capexp} and \eqref{powexp}]
\begin{align}
\tilde{C}_{iJ}(\ell) &:= \mathbb{E}\left[\sum_f C_{iJ}^f(\mathbf{p}(\mathbf{h};\ell),\mathbf{h})\right] \label{ctilde}\\
\tilde{P}_i(\ell) &:= \mathbb{E}\left[\sum_{f,J:(i,J)\in \mathcal{A}} p_{iJ}^f(\mathbf{h};\ell)\right].\label{ptilde}
\end{align}
These expectations can however be approximated by averaging over actual channel realizations. To do this, the power allocation subproblem \eqref{powcon} must be solved repeatedly for a prescribed number of time slots, say $S$, while using the same Lagrange multipliers. This would then allow approximation of the $\mathbb{E}$ operations in \eqref{ctilde} and \eqref{ptilde} with averaging operations, performed over channel realizations at these time slots.

It is evident however, that the averaging operation not only consumes $S$ time slots but also that the resulting subgradient is always outdated. Specifically, if the current time slot is of the form $\ell = KS +1$ with $K = 0, 1, 2, \ldots$, the most recent approximations of $\tilde{C}_{iJ}$ and $\tilde{P}_i$ available are
\begin{subequations}\label{hats}
\begin{align}
\hat{C}_{iJ}(\ell-S) &= \frac{1}{S}\sum_{\kappa = \ell-S}^{\ell-1}\sum_f C_{iJ}^f(\mathbf{p}(\mathbf{h}_\kappa;\ell-S),\mathbf{h}_\kappa) \label{chat}\\
\hat{P}_i(\ell-S) &= \frac{1}{S}\sum_{\kappa = \ell-S}^{\ell-1}\sum_{f,J:(i,J)\in \mathcal{A}} p_{iJ}^f(\mathbf{h}_\kappa;\ell-S).\label{phat}
\end{align}
\end{subequations}
Here, the power allocations are calculated using \eqref{powcon} with the \emph{old} multipliers $\lambda_{iJ}(\ell-S)$ and $\mu_{i}(\ell-S)$. The presence of outdated subgradient summands motivates the use of an asynchronous subgradient method such as the one in \cite{Kiwiel-Parallel}. 

Specifically, the dual updates still occur at every time slot but are allowed to use subgradients with outdated summands. Thus, $\hat{C}_{iJ}(\ell-S)$ and $\hat{P}_{i}(\ell-S)$ are used instead of the corresponding $\mathbb{E}[.]$ terms in \eqref{capexp} and \eqref{powexp} at the current time $\ell$. Further, since the averaging operation consumes another $S$ time slots, the same summands are also used for times $\ell+1$, $\ell+2$, $\ldots$, $\ell+S-1$. At time $\ell + S$, power allocations from the time slots $\ell$, $\ell+1$, $\ell+S-1$ become available, and are used for calculating $\hat{C}_{iJ}(\ell)$ and $\hat{P}_{i}(\ell)$, which then serve as the more recent subgradient summands. Note that a subgradient summand such as $\hat{C}_{iJ}$ is at least $S$ and at most $2S-1$ slots old. 

\begin{algorithm}[t]\label{innet}
\DontPrintSemicolon
\caption{Asynchronous Subgradient Algorithm}
Initialize $\boldsymbol{\zeta}(1) = 0$ and $\hat{C}_{iJ}(1) = \hat{P}_i(1) = 0$.\hspace{2cm}
Let $N$ be the maximum number of subgradient iterations.\nonumber\\
\For{$\ell =$ 1, 2, \ldots, $N$,}
{
Calculate primal iterates $a^m(\ell)$, $x_{ij}^{mt}(\ell)$, $z_{iJ}^m(\ell)$, $c_{iJ}(\ell)$, and $p_i(\ell)$ [cf.\ \eqref{rate}-\eqref{power}].\label{lineprimup}\\
Calculate the optimal power allocation $\mathbf{p}(\mathbf{h}_\ell;\tau(\ell))$ by solving \eqref{powcon} using $\mathbf{h}_\ell$ and $\boldsymbol{\zeta}(\tau(\ell))$.\\
Update dual iterates $\nu_i^{mt}(\ell+1)$, $\eta_{ik}^{mt}(\ell+1)$ and $\xi_{ij}(\ell+1)$ from the current primal iterates evaluated in Line \ref{lineprimup} [cf.\ \eqref{nut}-\eqref{xit}].\\
\If{$\ell - \tau(\ell) = S$,}
{
Calculate $\hat{C}_{iJ}(\tau(\ell))$ and $\hat{P}_{i}(\tau(\ell))$ as in \eqref{hats}.\label{lineexpec}\\
}
Update the dual iterates $\lambda_{iJ}(\ell+1)$ and $\mu_i(\ell+1)$: 
\begin{subequations}
\begin{align}
\lambda_{iJ}(\ell+1) &= \left[\lambda_{iJ}(\ell) + \epsilon(c_{iJ}(\ell) - \hat{C}_{iJ}(\tau(\ell)))\right]^+\nonumber\\
\mu_{i}(\ell+1) &= \left[\mu_{i}(\ell) + \epsilon(\hat{P}_i(\tau(\ell)) - p_i(\ell))\right]^+.\nonumber
\end{align}
\end{subequations}\\
Network Control: Use the current iterates $a^m(\ell)$ for flow control; $x_{ij}^{mt}(\ell)$ and $z_{iJ}^m(\ell)$ for routing and network coding; $c_{iJ}(\ell)$ for link rate control; and 
$\mathbf{p}(\mathbf{h}_\ell;\tau(\ell))$ for instantaneous power allocation.\\
}

\end{algorithm}

The asynchronous subgradient method is summarized as Algorithm \ref{innet}. The algorithm uses the function $\tau(\ell)$ which outputs the time of most recent averaging operation, that is,
\begin{equation}
\tau(\ell) = \max\bigl\{S\lfloor (\ell-S-1)/S\rfloor +1, 1\bigr\} ~~~\forall~~ \ell \geq 1.
\end{equation}
Note that $S \leq \ell - \tau(\ell) \leq 2S-1$. Recall also that the subgradient components $\hat{C}_{iJ}$ and $\hat{P}_{i}$ are evaluated only at times $\tau(\ell)$. 

The following proposition gives the dual convergence result on this algorithm. Define $\bar{G}$
as the bound $\left\|[\hat{\mathbf{C}}^T ~ \hat{\mathbf{P}}^T]^T\right\| \leq \bar{G}$ where $\hat{\mathbf{C}}$ and $\hat{\mathbf{P}}$ are formed by stacking the terms $\mathbb{E}\left[\sum_fC_{iJ}^f(\mathbf{p}(\mathbf{h}),\mathbf{h})\right]$ and $\mathbb{E}\left[\sum_{f,J}p_{iJ}^f(\mathbf{h})\right]$, respectively. 
\begin{prop}\label{dualasy}
If the maximum delay of the asynchronous counterparts of physical layer updates \eqref{lambdat} and \eqref{mut} is $D$, then:
\begin{enumerate}
	\item[a)] The sequence of dual iterates $\left\{\boldsymbol{\zeta}(\ell)\right\}$ is bounded; and
	\item[b)] The best dual value converges to $\mathsf{D}$ up to a constant:
	\begin{equation}
	\lim_{s\rightarrow \infty}\min_{1\leq \ell \leq s} \varrho(\boldsymbol{\zeta}(\ell)) \leq \mathsf{D} + 	\frac{\epsilon G^2}{2} + 2\epsilon D\bar{G}G.
	\label{asyncbestdualval}
	\end{equation}
\end{enumerate}
\end{prop}

Thus, the suboptimality in the asynchronous subgradient over the synchronous version is bounded by a constant proportional to $D = 2S-1$. Consequently, the asynchronous subgradient might need a smaller stepsize (and hence, more iterations) to reach a given distance from the optimal. 

The convergence of asynchronous subgradient methods for convex problems such as \eqref{dual} has been studied in \cite[Sec. 6]{Kiwiel-Parallel} for a diminishing stepsize. Proposition \ref{dualasy} provides a complementary result for constant stepsizes.  

Again, as with the synchronous version, the primal running averages also converge to within a constant from the optimal value of \eqref{netopt}. This is stated formally in the next proposition.
\begin{prop}\label{primalasy}
If the maximum delay of the asynchronous counterparts of physical layer updates \eqref{lambdat} and \eqref{mut} is $D$, then:
\begin{enumerate}
	\item[a)] There exists a sequence $\mathring{\mathbf{p}}(\mathbf{h};s)$ such that $(\bar{\mathbf{y}}(s),\mathring{\mathbf{p}}(\mathbf{h};s)) \in \mathcal{B}$ and 
\begin{equation}\label{asyfeas}
\lim_{s\rightarrow\infty}\left\|\left[\mathbf{q}(\bar{\mathbf{y}}(s),\mathring{\mathbf{p}}(\mathbf{h};s))\right]^{+}\right\| = 0.
\end{equation}
	\item[b)] The sequence $f(\bar{\mathbf{y}}(s))$ converges in the following sense:
	\vspace{-0.3cm}
\begin{subequations}\label{primaconst}
\begin{align}
\liminf_{s\rightarrow\infty} f(\bar{\mathbf{y}}(s)) &\geq \mathsf{P}-\frac{\epsilon G^2}{2} - 2\epsilon D\bar{G}G \label{inf-primaconst}\\
\text{and~~~~}\limsup_{s\rightarrow\infty} f(\bar{\mathbf{y}}(s)) &\leq \mathsf{P}. \label{sup-primaconst}
\end{align}
\end{subequations}
\end{enumerate}
\end{prop}

Note that as with the synchronous subgradient, the primal running averages are still asymptotically feasible, but the bound on their suboptimality increases by a term proportional to the delay $D$ in the physical layer updates. Of course, all the results in Propositions \ref{dualasy} and \ref{primalasy} reduce to the corresponding results in Propositions \ref{dualconv} and \ref{primconv} on setting $D = 0$. Interestingly, there is no similar result for primal convergence in asynchronous subgradient methods even for convex problems.

Finally, the following remarks on the online nature of the algorithm and the implementation of the Lagrangian maximizations in \eqref{updprime} are in order.
\begin{rem}
Algorithm 1 has several characteristics of an online adaptive algorithm. In particular, prior knowledge of the channel distribution is not needed in order to run the algorithm since the expectation operations are replaced by averaging over channel realizations on the fly. Likewise, running averages need not be evaluated; Proposition \ref{primalasy} ensures that the corresponding long-term averages will be near-optimal. Further, if at some time the network topology changes and the algorithm keeps running, it would be equivalent to restarting the entire algorithm with the current state as initialization. The algorithm is adaptive in this sense. 
\end{rem} 
\begin{rem}
Each of the maximization operations \eqref{rate}--\eqref{power} is easy, because it involves a single variable, concave objective, box constraints, and locally available Lagrange multipliers. The power control subproblem \eqref{powcon} however may be hard and require centralized computation in order to obtain a \mbox{(near-)} optimal solution. For the conflict graph model, see \cite{traskov,traskovd} and references therein for a list of approximate algorithms. For the SINR model, solutions of \eqref{powcon} could be based on approximation techniques in power control for digital subscriber lines (DSL)---see e.g., \cite{gatsis-journal} and references therein---and efficient message passing protocols as in \cite{xiyeh-ton}.
\end{rem}

\section{Numerical Tests}
\label{numerical}

\begin{figure}[tb]%
\centering
\includegraphics[scale = 0.65]{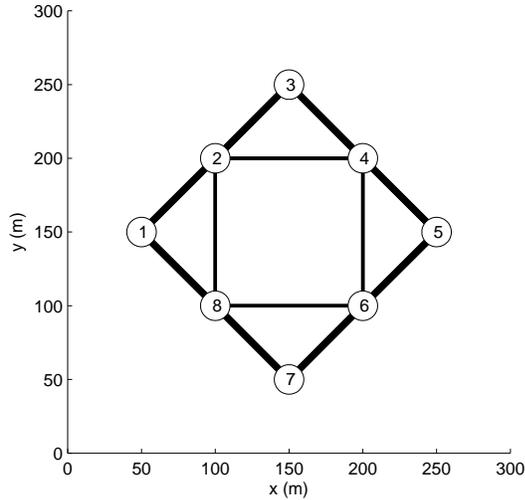}
\caption{The wireless network used in the simulations. The edges indicate the neighborhood of each node. The thickness of the edges is proportional to the mean of the corresponding channel.}%
\label{simnet}%
\end{figure}

The asynchronous algorithm developed in Section \ref{Sec:InNet} is simulated on the wireless network shown in Fig. \ref{simnet}. The network has 8 nodes placed on a 300m $\times$ 300m area. Hyperarcs originating from node $i$ are denoted by $(i,J) \in \mathcal{A}$ where $J \in 2^{N(i)}\setminus \emptyset$ i.e., the power set of the neighbors of $i$ excluding the empty set. For instance, hyperarcs originating from node 1 are $(1,\{2\})$, $(1,\{8\})$ and $(1,\{2,8\})$. The network supports the two multicast sessions $\mathcal{S}_1 = \{1,\{4,6\}\}$ and $\mathcal{S}_2 = \{4,\{1,7\}\}$. Table \ref{specs} lists the parameter values used in the simulation.  
\begin{table}[tb]
\renewcommand{\arraystretch}{1.2}
\caption{Simulation Parameters}
\centering
\resizebox{\figwidth}{!}{
\begin{tabular}{|c|l|}
\hline
$F$ & 2 \\
\hline
\multirow{4}{*}{$h_{ij}^f$} & Exponential with mean $\bar{h}_{ij}^f = 0.1(d_{ij}/d_0)^{-2}$ \\
& for all $(i,j)\in \mathcal{G}$ and $f$, where $d_0 = 20$m and \\ 
& $d_{ij}$  is the  distance between the nodes $i$ and $j$;\\
& links are reciprocal, i.e., $h_{ij}^f = h_{ji}^f$.\\
\hline
\multirow{2}{*}{$N_j$} & Noise power, evaluated using $d_{ij} = 100$m in the \\
& expression for $\bar{h}_{ij}^f$ above\\
\hline
$p_{\text{max}}^f$ & $5$ W/Hz for all $f$\\
\hline
$p^{\text{max}}_{i}$ & $5$ W/Hz for all $i\in\mathcal{N}$\\
\hline
$a_{\text{max}}^m$ & $5$ bps/Hz for all $m$\\
\hline
$a_{\text{min}}^m$ & $10^{-4}$ bps/Hz for all $m$\\
\hline
\multirow{2}{*}{$c_{iJ}^{\text{max}}$} & interference-free capacity obtained for each $j\in J$ via  \\
& waterfilling under $\mathbb{E}\left[\sum_fp^f(h_{ij}^f)\right] \leq p^{\text{max}}_i$ for all $i\in\mathcal{N}$\\
\hline
$z^{\text{max}}_{iJ}$ & $c^{\text{max}}_{iJ}/2$ for all $(i,J)\in\mathcal{A}$ \\
\hline
$x^{\text{max}}_{ij}$ & $z_{iJ}^{\text{max}}/2$ for $j\in J$ and $i\in\mathcal{N}$ \\
\hline
$U_m(a^m)$ & $\ln(a^m)$ for all $m$\\
\hline
$V_i(p_i)$ & $10p_i^2$ for all $i\in\mathcal{N}$\\
\hline
\end{tabular}}
\label{specs}
\vspace{.1cm}
\end{table}

The conflict graph model of Example 1 with secondary interference constraints is used. In order to solve the power control subproblem \eqref{powcon}, we need to enumerate all possible sets of conflict free hyperarcs (cf.\ Example 1); these sets are called matchings. At each time slot, the aim is to find the matching that maximizes the objective function $\sum_{f,(i,J)}\gamma_{iJ}^f$. Note that since $\gamma_{iJ}^f$ is a positive quantity, only maximal matchings, i.e., matchings with maximum possible cardinality, need to be considered. At each time slot, the following two steps are carried out.
\begin{enumerate}
	\item[S1)] Find the optimal power allocation for each maximal matching. Note that the capacity of an active hyperarc is a function of the power allocation over that hyperarc alone [cf.\ \eqref{snr} and \eqref{harccap}]. Thus, the maximization in \eqref{powcon} can be solved separately for each hyperarc and tone. The resulting objective [cf.\ \eqref{gamma}] is a concave function in a single variable, admitting an easy waterfilling-type solution.
	\item[S2)] Evaluate the objective function \eqref{powcon} for each maximal matching and for powers found in Step 2, and choose the matching with the highest resulting objective value.
\end{enumerate}

It is well known that the enumeration of hyperarc matchings requires exponential complexity \cite{traskov}. Since the problem at hand is small, full enumeration is used. 

\begin{figure}[tb]
\centering
\includegraphics[width=0.85\figwidth]{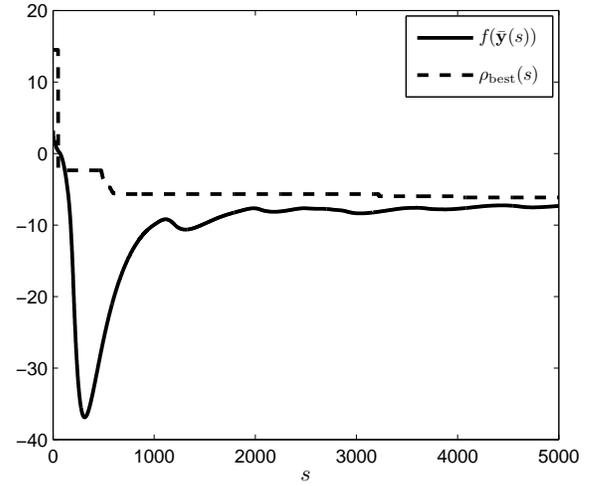}
\caption{Evolution of the utility function $f(\bar{\mathbf{y}}(s))$ and best dual value $\rho_{\mathrm{best}}(s) = \min_{\ell\leq s} \varrho(\bm\zeta(\ell))$ for $\epsilon = 0.15$ and $S = 50$.}%
\label{zerodualfig}%
\end{figure}
Fig. \ref{zerodualfig} shows the evolution of the utility function $f(\bar{\mathbf{y}}(s))$  and the best dual value up to the current iteration. The utility function is evaluated using the running average of the primal iterates [cf. \eqref{runav}]. It can be seen that after a certain number of iterations, the primal and dual values remain very close corroborating the vanishing duality gap.

\begin{figure}[tb]
\centering
\includegraphics[width = 0.91\figwidth]{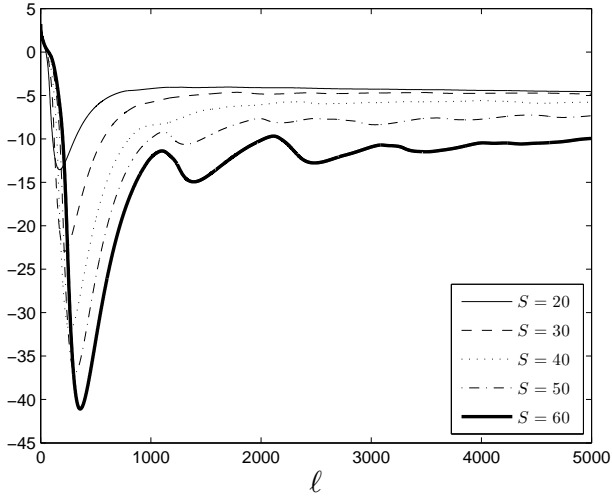}
\caption{Evolution of the utility function $f(\bar{\mathbf{y}}(s))$ for different values of $S$ with stepsize $\epsilon = 0.15$.}%
\label{utileval}%
\end{figure}
Fig. \ref{utileval} shows the evolution of the utility function for different values of $S$. Again the utility function converges to a near-optimal value after sufficient number of iterations. Note however that the gap from the optimal dual value increases for large values of $S$, such as $S = 60$ (cf.\ Proposition \ref{primalasy}). 

\begin{figure}[tb]
\centering
\includegraphics[width = 0.9\figwidth]{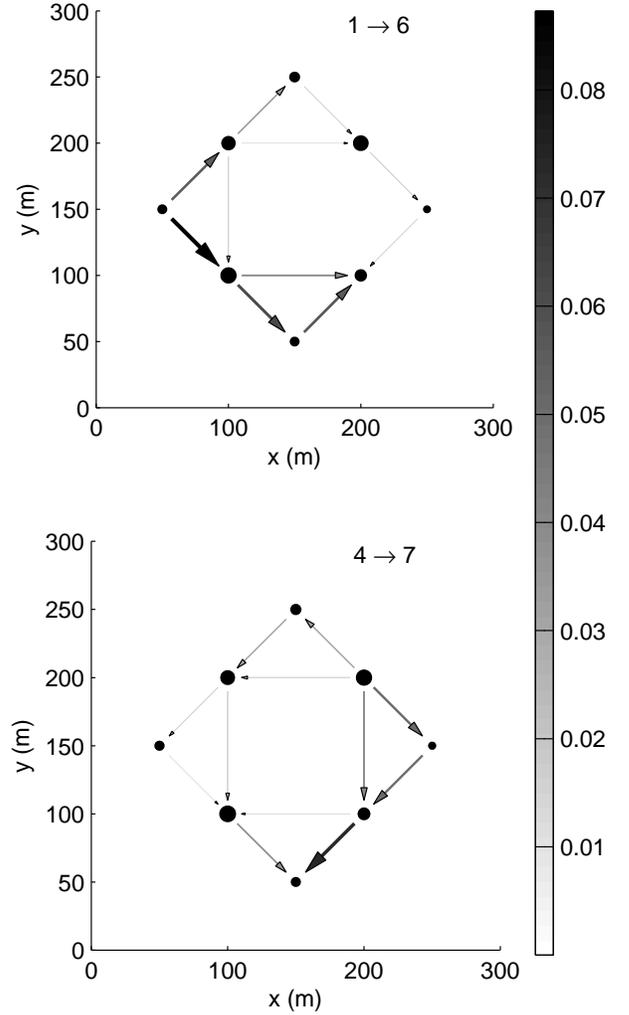}
\caption{Some of the optimal primal values after 5000 iterations with $\epsilon = 0.15$ and $S = 40$. The gray level of the edges corresponds to values of virtual flows according to the color bar on the right, with units bps/Hz.}%
\label{drawall}
\end{figure}
Finally, Fig. \ref{drawall} shows the optimal values of certain optimization variables. Specifically the two subplots show all the virtual flows to given sinks for each of the multicast sessions, namely, $\{s^1 =1,  t = 6\}$ and $\{s^2 = 4, t = 7\}$, respectively. The thickness and the gray level of the edges is proportional to the magnitude of the virtual flows. It can be observed that most virtual flows are concentrated along the shorter paths between the source and the sink. Also, the radius of the circles representing the nodes is proportional to the optimal average power consumption. It can be seen that the inner nodes 2, 4, 6, and 8 consume more power than the outer ones, 1, 3, 5, and 7. This is because the inner nodes have more neighbors, and thus more opportunities to transmit. Moreover, the outer nodes are all close to their neighbors.

\section{Conclusions}\label{concl}
This paper formulates a cross-layer optimization problem for multicast networks where nodes perform intra-session network coding, and operate over fading broadcast links. Zero duality gap is established, rendering layered architectures optimal.

Leveraging this result, an adaptation of the subgradient method suitable for network control is also developed. The method is asynchronous, because the physical layer returns its contribution to the subgradient vector with delay. Using the subgradient vector, primal iterates in turn dictate routing, network coding, and resource allocation. It is established that network variables, such as the long-term average rates admitted into the network layer, converge to near-optimal values, and the suboptimality bound is provided explicitly as a function of the delay in the subgradient evaluation. 

\appendices

\section{Strong Duality for the Networking Problem \eqref{netopt}}\label{zerodual}

This appendix formulates a general version of problem \eqref{netopt}, and gives results about its duality gap. Let $\bfh$ be the random channel vector in $\Omega:=\mathds{R}_+^{d_\bfh}$, where $\mathds{R}_+$ denotes the nonnegative reals, and $d_\bfh$ the dimensionality of $\bfh$. Let $\mathcal{D}$ be the $\sigma$-field of Borel sets in $\Omega$, and $P_\bfh$ the distribution of $\bfh$, which is a probability measure on $\mathcal{D}$.

As in \eqref{netopt}, consider two optimization variables: the vector $\av$ constrained to a subset $\boxv$ of the Euclidean space $\mathds{R}^{d_\av}$; and the function $\bfp:\Omega \rightarrow \mathds{R}^{d_\bfp}$ belonging to an appropriate set of functions $\setp$. In the networking problem, the aforementioned function is the power allocation $\bfp(\bfh)$, and set $\setp$ consists of the power allocation functions satisfying instantaneous constraints, such as spectral mask or hyperarc scheduling constraints (cf. also Examples 1 and 2). Henceforth, the function variable will be denoted by $\bfp$ instead of $\bfp(\bfh)$, for brevity. Let $\Pi$ be a subset of $\mathds{R}^{d_\bfp}$. Then $\setp$ is defined as the set of functions taking values in $\Pi$.
\begin{equation}
\setp:=\{\bfp\;\text{measurable}\,| \, \bfp(\bfh)\in \Pi \:\text{for almost all}\: \bfh\in\Omega \}.
\end{equation}

The network optimization problem \eqref{netopt} can be written in the general form
\begin{subequations}
\label{Eq:App-OptPr-All}
\begin{IEEEeqnarray}{r'r'l}
\mathsf{P} =&\max & f(\av) \label{Eq:App-OptPr-obj}\\
&\text{subj. to} & \gx(\av)+\exv[\gp(\bfp(\bfh),\bfh)] \leq \bm{0} \label{Eq:App-OptPr-IneqConstr}\\
& & \av\in\boxv, \: \bfp \in\setp \label{Eq:App-OptPr-BoxConstr}
\end{IEEEeqnarray}
\end{subequations}
where $\gx$ and $\gp$ are $\mathds{R}^d$-valued functions describing $d$ constraints. The formulation also subsumes similar problems in the unicast routing framework such as those in \cite{ale09, gatsis-journal}. 

Evidently, problem \eqref{netopt} is a special case of \eqref{Eq:App-OptPr-All}. If inequalities \eqref{flowconr}--\eqref{powr} are rearranged to have zeros on the right hand side, function $\gp(\bfp(\bfh),\bfh)$ will simply have zeros in the entries that correspond to constraints \eqref{flowconr}--\eqref{arccapr}. The function $\mathbf{q}(\mathbf{y},\mathbf{p}(\mathbf{h}))$ defined before \eqref{lagrangian} equals $\gx(\av)+\exv[\gp(\bfp(\bfh),\bfh)]$. 

The following assumptions regarding \eqref{Eq:App-OptPr-All} are made:
\begin{assumption}
\label{As:box}
Constraint set $\boxv$ is convex, closed, bounded, and in the interior of the domains of functions $f(\av)$ and $\gx(\av)$. Set $\Pi$ is closed, bounded, and in the interior of the domain of function $\gp(.,\bfh)$ for all \bfh. 
\end{assumption}
\begin{assumption}
\label{As:convexity}
Function $f(\cdot)$ is concave, $\gx(\cdot)$ is convex, and $\gp(\bfp(\bfh),\bfh)$ is integrable whenever $\bfp$ is measurable. Furthermore, there is a $\bar G>0$ such that $\left\|\exv[\gp(\bfp(\bfh),\bfh)]\right\|\leq \bar G$, whenever $\bfp\in\setp$.
\end{assumption}
\begin{assumption}
\label{As:contcdf}
Random vector $\bfh$ is continuous;\footnote{Formally, this is equivalent to saying that $P_\bfh$ is absolutely continuous with respect to the Lebesgue measure on $\mathds{R}_+^{d_\bfh}$. In more practical terms, it means that $\bfh$ has a probability density function without deltas.} and
\end{assumption}
\begin{assumption}
\label{As:Slater}
There exist $\av'\in\boxv$ and $\bfp'\in\setp$ such that \eqref{Eq:App-OptPr-IneqConstr} holds as strict inequality (Slater constraint qualification).
\end{assumption}
Note that these assumptions are natural for the network optimization problem \eqref{netopt}. Specifically, $\mathcal{B}_\av$ are the box constraints for variables $a^m$, $x_{ij}^{mt}$, $z_{iJ}^m$, $c_{iJ}$, and $p_i$; and $\Pi$ gives the instantaneous power allocation constraints. The function $f(\mathbf{y})$ is selected concave and $g(\mathbf{y})$ is linear. Moreover, the entries of $\gp(\bfp(\bfh),\bfh)$ corresponding to \eqref{powr} are bounded  because the set $\Pi$ is bounded. 
For the same reason, the ergodic capacities $\mathbb{E}[C_{iJ}^f(\mathbf{p}(\mathbf{h}),\mathbf{h})]$ are bounded.

While \eqref{Eq:App-OptPr-All} is not convex in general, it is separable \cite[Sec. 5.1.6]{Bertsekas-NonlinProg}. The Lagrangian (keeping constraints \eqref{Eq:App-OptPr-BoxConstr} implicit) and the dual function are, respectively [cf. also \eqref{lagrangian} and \eqref{dualf}]
\begin{align}
\mathcal{L}(\av,\bfp,\bm\zeta)&=f(\av)-\bm\zeta^T \Big(\gx(\av)+\exv[\gp(\bfp(\bfh),\bfh)] \Big)\label{Eq:App-LagrFcn}\\
\varrho(\bm\zeta)&:=\max_{\av\in\boxv,\,\bfp\in\setp} \mathcal{L}(\av,\bfp,\bm\zeta) = \psi(\bm\zeta)+\phi(\bm\zeta).\label{Eq:App-DualFcn}
\end{align}
where $\bm\zeta$ denotes the vector of Lagrange multipliers and
\begin{subequations}
\begin{align}
\psi(\bm\zeta)&:=\max_{\av\in\boxv} \left\{f(\av)-\bm\zeta^T\gx(\av)\right\}\\
\phi(\bm\zeta)&:=\max_{\bfp\in\setp}\bm\zeta^T\exv[\gp(\bfp(\bfh),\bfh).
\end{align}
\end{subequations}
The additive form of the dual function is a consequence of the separable structure of the Lagrangian. Further, AS\ref{As:box} and AS\ref{As:convexity} ensure that the domain of $\varrho(\bm\zeta)$ is $\mathds{R}^d$. Finally, the dual problem becomes [cf. also \eqref{dual}]
\begin{equation}
\label{Eq:App-DualPr}
\mathsf{D} = \min_{\bm\zeta \geq \bm{0}} \varrho(\bm\zeta).
\end{equation}

As $\bfp$ varies in $\setp$, define the range of $\exv[\gp(\bfp(\bfh),\bfh)]$  as
\begin{IEEEeqnarray}{c}
\label{Eq:App-RangeDef}
\mathcal{R}:=\left\{\mathbf{w}\in\mathds{R}^d \left| \mathbf{w}=\exv[\gp(\bfp(\bfh),\bfh)] \text{~for some~} \bfp\in\setp \right. \right\}.\IEEEeqnarraynumspace
\end{IEEEeqnarray}
The following lemma demonstrating the convexity of $\mathcal{R}$ plays a central role in establishing the zero duality gap of \eqref{Eq:App-OptPr-All}, and in the recovery of primal variables from the subgradient method.
\begin{lem}
\label{Lem:ConvexRange}
If AS\ref{As:box}-AS\ref{As:contcdf} hold, then the set $\mathcal{R}$ is convex.
\end{lem}

The proof relies on Lyapunov's convexity theorem \cite{Blackwell-OnLyapunov}. Recently, an extension of Lyapunov's theorem \cite[Extension 1]{Blackwell-OnLyapunov} has been applied to show zero duality gap of power control problems in DSL \cite{Luo-Zhang-DSM}. This extension however does not apply here, as indicated in the ensuing proof. 
In a related contribution \cite{ale09}, it is shown that the perturbation function of a problem similar to \eqref{Eq:App-OptPr-All} is convex; the claim of Lemma~\ref{Lem:ConvexRange} though is quite different.

\begin{IEEEproof}[Proof of Lemma~\ref{Lem:ConvexRange}]
Let $\mathbf{r}_1$ and $\mathbf{r}_2$ denote arbitrary points in $\mathcal{R}$, and let $\alpha\in(0,1)$ be arbitrary. By the definition of $\mathcal{R}$, there are functions $\bfp_1$ and $\bfp_2$ in $\setp$ such that 
\begin{equation}
\label{Eq:App-rpoints}
\mathbf{r}_1=\int \gp(\bfp_1(\bfh),\bfh)dP_\bfh~\text{and}~\mathbf{r}_2=\int \gp(\bfp_2(\bfh),\bfh)dP_\bfh.
\end{equation}
Now define
\begin{equation}
\mathbf{u}(E):=\left[\begin{IEEEeqnarraybox}[][c]{,c,}
\int_E \gp(\bfp_1(\bfh),\bfh)dP_\bfh\\
\int_E \gp(\bfp_2(\bfh),\bfh)dP_\bfh%
\end{IEEEeqnarraybox}\right], \quad E\in\mathcal{D}.
\end{equation}
The set function $\mathbf{u}(E)$ is a nonatomic vector measure on $\mathcal{D}$, because $P_\bfh$ is nonatomic (cf. AS\ref{As:contcdf}) and the functions $\gp(\bfp_1(\bfh),\bfh)$ and $\gp(\bfp_2(\bfh),\bfh)$ are integrable (cf. AS\ref{As:convexity}); see \cite{Dinculeanu-VecMeasures} for definitions. Hence, Lyapunov's theorem applies to $\mathbf{u}(E)$; see also \cite[Extension 1]{Blackwell-OnLyapunov} and \cite[Lemma~1]{ale09}. 

Specifically, consider a null set $\Phi$ in $\mathcal{D}$, i.e., a set with $P_\bfh(\Phi)=0$, and the whole space $\Omega\in\mathcal{D}$. It holds that $\mathbf{u}(\Phi)=\bm{0}$ and $\mathbf{u}(\Omega)=[\mathbf{r}_1^T,\mathbf{r}_2^T]^T$. For the chosen $\alpha$, Lyapunov's theorem asserts that there exists a set $E_\alpha\in\mathcal{D}$ such that ($E_\alpha^c$ denotes the complement of $E_\alpha$)
\begin{subequations}
\label{Eq:App-vecmeasE}
\begin{IEEEeqnarray}{rCl}
\mathbf{u}(E_\alpha) & = &\alpha \mathbf{u}(\Omega) +(1-\alpha) \mathbf{u}(\Phi)= \alpha
\left[\begin{IEEEeqnarraybox}[][c]{,c,}
\mathbf{r}_1\\\mathbf{r}_2 \end{IEEEeqnarraybox}\right] \\
\mathbf{u}(E_\alpha^c) & = & \mathbf{u}(\Omega) - \mathbf{u}(E_\alpha) = 
(1-\alpha) \left[\begin{IEEEeqnarraybox}[][c]{,c,}
\mathbf{r}_1\\\mathbf{r}_2 \end{IEEEeqnarraybox}\right].
\end{IEEEeqnarray}
\end{subequations}

Now using these $E_\alpha$ and $E_\alpha^c$, define
\begin{equation}
\bfp_{\alpha}(\bfh)=\begin{cases}
\bfp_1(\bfh),& \bfh\in E_\alpha\\
\bfp_2(\bfh),& \bfh\in E_\alpha^c.
\end{cases}
\end{equation}
It is easy to show that $\bfp_{\alpha}(\bfh)\in\setp$. In particular, the function $\bfp_{\alpha}(\bfh)$ can written as $\bfp_{\alpha}(\bfh)=\bfp_1(\bfh)\ind_{E_\alpha}+\bfp_2(\bfh)\ind_{E_\alpha^c}$, where $\ind_{E}$ is the indicator function of a set $E\in\mathcal{D}$. Hence it is measurable, as sum of measurable functions. 
Moreover, we have that $\bfp_{\alpha}(\bfh)\in\Pi$ for almost all $\bfh$, because $\bfp_1(\bfh)$ and $\bfp_2(\bfh)$ satisfy this property.  The need to show $\bfp_{\alpha}(\bfh)\in\setp$ makes \cite[Extension 1]{Blackwell-OnLyapunov} not directly applicable here.

Thus, $\bfp_{\alpha}(\bfh)\in\setp$ and satisfies [cf.~\eqref{Eq:App-vecmeasE}]
\begin{multline}
\int \gp(\bfp_{\alpha}(\bfh),\bfh)dP_\bfh = 
\int_{E_\alpha} \gp(\bfp_{1}(\bfh),\bfh)dP_\bfh\\
  {+}\: \int_{E_\alpha^c} \gp(\bfp_{2}(\bfh),\bfh)dP_\bfh = 
  \alpha \mathbf{r}_1 +(1-\alpha) \mathbf{r}_2.
\end{multline} 
Therefore, $\alpha \mathbf{r}_1 +(1-\alpha) \mathbf{r}_2 \in\mathcal{R}$.
\end{IEEEproof}

Finally, the zero duality gap result follows from Lemma \ref{Lem:ConvexRange} and is stated in the following proposition.
\begin{prop}
\label{Prop:ZeroGap}
If AS\ref{As:box}-AS\ref{As:Slater} hold, then problem \eqref{Eq:App-OptPr-All} has zero duality gap, i.e.,
$\mathsf{P}=\mathsf{D}$.
Furthermore, the values $\mathsf{P}$ and $\mathsf{D}$ are finite, the dual problem \eqref{Eq:App-DualPr} has an optimal solution, and the set of optimal solutions of \eqref{Eq:App-DualPr} is bounded.
\end{prop}

\begin{IEEEproof}
Function $f(\av)$ is continuous on $\boxv$ since it is convex (cf. AS\ref{As:box} and AS\ref{As:convexity}) \cite[Prop.~1.4.6]{Bertsekas-ConvexAnalysis}. This, combined with the compactness of $\boxv$, shows that the optimal primal value $\mathsf{P}$ is finite.
Consider the set
\begin{align}
\mathcal{W} := & \left\{  (w_1,\ldots,w_d,u)\in \mathds{R}^{d+1} ~ \left|~ \:f(\av)\leq u, \right.\right. \nonumber\\
& \left. \gx(\av) + \exv[\gp(\bfp(\bfh),\bfh)] \leq \mathbf{w} \text{~for some~} \av\in\boxv, \,\bfp\in\setp \right\}. 
\label{Eq:App-episet}
\end{align}

Using Lemma \ref{Lem:ConvexRange}, it is easy to verify that set $\mathcal{W}$ is convex. The rest of the proof follows that of \cite[Prop.~5.3.1 and 5.1.4]{Bertsekas-NonlinProg}, using the finiteness of $\mathsf{P}$ and Slater constraint qualification (cf. AS\ref{As:Slater}).

The boundedness of the optimal dual set is a standard result for convex problems under Slater constraint qualification and finiteness of optimal primal value; see e.g., \cite[Prop. 6.4.3]{Bertsekas-ConvexAnalysis} and \cite[p. 1762]{nedic07}. 
The proof holds also in the present setup since $\mathsf{P}$ is finite, $\mathsf{P}=\mathsf{D}$, and AS\ref{As:Slater} holds.
\end{IEEEproof}

\section{Dual and Primal Convergence Results} \label{subapp}
This appendix formulates the synchronous and asynchronous subgradient methods for the generic problem~\eqref{Eq:App-OptPr-All}; and establishes the convergence claims in Propositions \ref{dualconv}--\ref{primalasy}. Note that Propositions \ref{dualconv} and \ref{primconv} follow from Propositions \ref{dualasy} and \ref{primalasy}, respectively, upon setting the delay $D = 0$. 

Starting from an arbitrary $\bm\zeta(1)\geq\bm{0}$, the subgradient iterations for \eqref{Eq:App-DualPr} indexed by $\ell\in\mathds{N}$ are [cf. also \eqref{updprdu}]
\begin{subequations}
\label{Eq:App-MaxSeq-ALL}
\begin{align}
\av(\ell) 				& \in  \argmax_{\av\in\boxv} \,\left\{f(\av)-\bm\zeta^T(\ell)\gx(\av)\right\} \label{Eq:App-MaxSeq-v}\\
\bfp(.;\ell) 	& \in  \argmax_{\bfp\in\setp}\,\bm\zeta^T(\ell)\exv[\gp(\bfp(\bfh),\bfh)] \label{Eq:App-MaxSeq-ph}\\
\text{and}\hspace{1cm}
\bm\zeta(\ell+1) &= \left[\bm\zeta(\ell)+\epsilon\left(\sg(\ell)+\sph(\ell)\right)\right]^+\label{Eq:App-SyncSgUpdate}
\end{align}
\end{subequations}
where $\check{\mathbf{g}}(\ell)$ and $\check{\mathbf{v}}(\ell)$ are the subgradients of functions $\psi(\bm\zeta)$ and $\phi(\bm\zeta)$, defined as [cf. also \eqref{subgrads}]
\begin{subequations}
\label{Eq:App-SubgradDef-ALL}
\begin{align}
\sg(\ell) & :=  \gx(\av(\ell)) \label{Eq:App-SubgradDef-y}\\
\sph(\ell) & :=  \exv[\gp(\bfp(\bfh;\ell),\bfh)]. \label{Eq:App-SubgradDef-ph}
\end{align}
\end{subequations}
The iteration in \eqref{Eq:App-SyncSgUpdate} is synchronous, because at every $\ell$, both maximizations \eqref{Eq:App-MaxSeq-v} and \eqref{Eq:App-MaxSeq-ph} are performed using the current Lagrange multiplier $\bm\zeta(\ell)$. An \emph{asynchronous} method is also of interest and operates as follows. Here, the component $\sph$ of the overall subgradient used at $\ell$ does not necessarily correspond to the Lagrange multiplier $\bm\zeta(\ell)$, but to the Lagrange multiplier at a time $\tau(\ell)\leq \ell$. Noting that the maximizer in \eqref{Eq:App-MaxSeq-ph} is $\bfp(.;\tau(\ell)))$ and the corresponding subgradient component used at $\ell$ is $\sph(\tau(\ell))$, the iteration takes the form
\begin{equation}
\label{Eq:App-AsyncSgUpdate}
\bm\zeta(\ell+1) = \left[\bm\zeta(\ell)+\epsilon\left(\sg(\ell)+\sph(\tau(\ell))\right)\right]^+, \: \ell\in\mathds{N}.
\end{equation}
The difference $\ell-\tau(\ell)$ is the delay with which the subgradient component $\sph$ becomes available. In Algorithm 1 for example, the delayed components are $\hat{C}_{iJ}(\tau(\ell))$ and $\hat{P}_i(\tau(\ell))$.

Next, we proceed to analyze the convergence of \eqref{Eq:App-AsyncSgUpdate}. Function $\gx(\av)$ is continuous on $\boxv$ because it is convex \cite[Prop.~1.4.6]{Bertsekas-ConvexAnalysis}. Then AS\ref{As:box} and AS\ref{As:convexity} imply that there exists a bound $G$ such that for all $\av\in\boxv$ and $\bfp\in\setp$,
\begin{equation}
\label{Eq:App-SubgradNormBd}
\left\|\gx(\av) + \exv[\gp(\bfp(\bfh),\bfh)]\right\|\leq G.
\end{equation}

Due to this bound on the subgradient norm, algorithm \eqref{Eq:App-AsyncSgUpdate} can be viewed as a special case of an approximate subgradient method \cite{Nedic-DeterminNoiseSubgrad}. We do not follow this line of analysis here though, because it does not take advantage of the source of the error in the subgradient---namely, that an old maximizer of the Lagrangian is used.  Moreover, algorithm \eqref{Eq:App-AsyncSgUpdate} can be viewed as a particular case of an $\varepsilon$-subgradient method (see \cite[Sec. 6.3.2]{Bertsekas-NonlinProg} for definitions). This connection is made in \cite{Kiwiel-Parallel} which only deals with diminishing stepsizes; here results are proved for constant stepsizes. The following assumption is adopted for the delay $\ell-\tau(\ell)$.
\begin{assumption}
\label{As:BddDelay}
There exists a finite $D\in \mathds{N}$ such that $\ell-\tau(\ell)\leq D$ for all $\ell\in\mathds{N}$.
\end{assumption}
AS\ref{As:BddDelay} holds for Algorithm 1 since the maximum delay there is $D = 2S-1$. The following lemma collects the results needed for Propositions \ref{dualconv} and \ref{dualasy}. Specifically, it characterizes the error term in the subgradient definition when $-\sph(\tau(\ell))$ is used; and also relates successive iterates $\bm\zeta(\ell)$ and $\bm\zeta(\ell+1)$. The quantity $\bar G$ in the ensuing statement was defined in AS\ref{As:convexity}.

\begin{lem}
\label{Lem:DualIterRel}
Under AS\ref{As:box}-AS\ref{As:BddDelay}, the following hold for the sequence $\{\bm\zeta(\ell)\}$ generated by \eqref{Eq:App-AsyncSgUpdate} for all $\bm\theta\geq \bm{0}$
\begin{subequations}
\begin{align}
&\textnormal{a)}~ -\sph^T(\tau(\ell))\left(\bm\theta-\bm\zeta(\ell) \right)  \nonumber\\ 
& \hspace{3cm}\leq \phi(\bm\theta)-\phi(\bm\zeta(\ell))+2\epsilon D G \bar G \label{Eq:App-epssg-v}\\
&\textnormal{b)}~ -\left(\sg(\ell)+\sph(\tau(\ell))\right)^T \left(\bm\theta-\bm\zeta(\ell) \right) \nonumber\\
& \hspace{3cm}\leq   \varrho(\bm\theta)-\varrho(\bm\zeta(\ell))+2\epsilon D G \bar G \label{Eq:App-epssg-all}\\
&\textnormal{c)}~ \left\|\bm\zeta(\ell+1)-\bm\theta\right\|^2 - \left\|\bm\zeta(\ell)-\bm\theta\right\|^2 \nonumber\\
&\hspace{1.2cm} \leq  2\epsilon \left[\varrho(\bm\theta)-\varrho(\bm\zeta(\ell)) \right] + \epsilon^2 {G}^2+ 4 \epsilon^2 DG\bar G \label{Eq:App-basiter}
\end{align}
\end{subequations}
\end{lem}
Parts (a) and (b) of Lemma \ref{Lem:DualIterRel} assert that the vectors $-\sph(\tau(\ell))$ and $-\sg(\ell)-\sph(\tau(\ell))$ are respectively $\varepsilon$-subgradients of $\phi(\bm\zeta)$ and the dual function $\varrho(\bm\zeta)$ at $\bm\zeta(\ell)$, with $\varepsilon=2\epsilon D G \bar G$.  Note that $\varepsilon$ is a constant proportional to the delay $D$. 

\begin{IEEEproof}[Proof of Lemma \ref{Lem:DualIterRel}]
a) 
The left-hand side of \eqref{Eq:App-epssg-v} is
\begin{IEEEeqnarray}{rCl}
-\sph^T(\tau(\ell))\left(\bm\theta-\bm\zeta(\ell) \right) & = & 
-\sph^T(\tau(\ell))\left[\bm\theta-\bm\zeta(\tau(\ell)) \right] \nonumber\\
 & & -\:\sph^T(\tau(\ell))\left[\bm\zeta(\tau(\ell))-\bm\zeta(\ell) \right].
 \IEEEeqnarraynumspace
\label{Eq:Pf-epssg-v-1}
\end{IEEEeqnarray}
Applying the definition of the subgradient for $\phi(\bm\zeta)$ at $\bm\zeta(\tau(\ell))$ to \eqref{Eq:Pf-epssg-v-1}, it follows that
\begin{IEEEeqnarray}{rCl}
-\sph^T(\tau(\ell))\left(\bm\theta-\bm\zeta(\ell) \right) & \leq & 
\phi(\bm\theta)-\phi(\bm\zeta(\tau(\ell))) \nonumber \\
& & -\:\sph^T(\tau(\ell))\left[\bm\zeta(\tau(\ell))-\bm\zeta(\ell) \right].
\IEEEeqnarraynumspace\label{Eq:Pf-epssg-v-2}
\end{IEEEeqnarray}

Now, adding and subtracting the same terms in the right-hand side of \eqref{Eq:Pf-epssg-v-2}, we obtain
\begin{multline}
-\sph^T(\tau(\ell))\left(\bm\theta-\bm\zeta(\ell) \right) \leq 
\phi(\bm\theta)-\phi(\bm\zeta(\ell)) \\
+\sum_{\kappa=1}^{\ell-\tau(\ell)} \left[\phi\bigl(\bm\zeta(\tau(\ell)+\kappa) \bigr) - \phi\bigl(\bm\zeta(\tau(\ell)+\kappa-1)\bigr) \right] \\
-\sum_{\kappa=1}^{\ell-\tau(\ell)} \sph^T(\tau(\ell)) \left[\bm\zeta(\tau(\ell)+\kappa - 1) - \bm\zeta(\tau(\ell)+\kappa)\right].
\label{Eq:Pf-epssg-v-3}
\end{multline}
Applying the definition of the subgradient for $\phi(\bm\zeta)$ at $\bm\zeta(\tau(\ell)+\kappa)$ to \eqref{Eq:Pf-epssg-v-3}, it follows that
\begin{multline}
-\sph^T(\tau(\ell))\left(\bm\theta-\bm\zeta(\ell) \right) \leq 
\phi(\bm\theta)-\phi(\bm\zeta(\ell)) \\
+\sum_{\kappa=1}^{\ell-\tau(\ell)} \sph^T(\tau(\ell)+\kappa) \left[\bm\zeta(\tau(\ell)+\kappa - 1) - \bm\zeta(\tau(\ell)+\kappa)\right] \\
-\sum_{\kappa=1}^{\ell-\tau(\ell)} \sph^T(\tau(\ell)) \left[\bm\zeta(\tau(\ell)+\kappa - 1) - \bm\zeta(\tau(\ell)+\kappa)\right].
\label{Eq:Pf-epssg-v-4}
\end{multline}
Using the Cauchy-Schwartz inqeuality, \eqref{Eq:Pf-epssg-v-4} becomes
\begin{multline}
-\sph^T(\tau(\ell))\left(\bm\theta-\bm\zeta(\ell) \right) \leq 
\phi(\bm\theta)-\phi(\bm\zeta(\ell)) \\
+\sum_{\kappa=1}^{\ell-\tau(\ell)} \left(\left\|\sph(\tau(\ell)+\kappa)\right\| + \left\|\sph(\tau(\ell))\right\| \right)\\ \cdot \left\| \bm\zeta(\tau(\ell)+\kappa - 1) - \bm\zeta(\tau(\ell)+\kappa)\right\|.
\label{Eq:Pf-epssg-v-5}
\end{multline}

Now, write the subgradient iteration [cf. \eqref{Eq:App-AsyncSgUpdate}] at $\tau(\ell)+\kappa - 1$:
\begin{multline}
\label{Eq:Pf-updtimekappa}
\bm\zeta(\tau(\ell)+\kappa) = \bigl[\bm\zeta(\tau(\ell)+\kappa-1)\\+{\:}\epsilon\bigl(\sg(\tau(\ell)+\kappa-1)+
\sph(\tau(\tau(\ell)+\kappa-1))\bigr)\bigr]^+.
\end{multline}
Subtracting $\bm\zeta(\tau(\ell)+\kappa-1)$ from both sides of the latter and using the nonexpansive property of the projection~\cite[Prop.~2.2.1]{Bertsekas-ConvexAnalysis} followed by \eqref{Eq:App-SubgradNormBd}, one finds from \eqref{Eq:Pf-updtimekappa} that
\begin{multline}
\label{Eq:Pf-difzetabd}
||\bm\zeta(\tau(\ell)+\kappa) - \bm\zeta(\tau(\ell)+\kappa-1)|| \\ \leq \epsilon\left\|\sg(\tau(\ell)+\kappa-1)+
\sph(\tau(\tau(\ell)+\kappa-1))\right\| \leq \epsilon G.
\end{multline}

Finally, recall that $\left\|\sph(\ell)\right\|\leq \bar G$ for all $\ell\in\mathds{N}$ (cf. AS\ref{As:convexity}), and $\ell - \tau(\ell)\leq D$ for all $\ell\in\mathds{N}$ (cf. AS\ref{As:BddDelay}). Applying the two aforementioned assumptions and \eqref{Eq:Pf-difzetabd} to \eqref{Eq:Pf-epssg-v-5}, we obtain \eqref{Eq:App-epssg-v}.

b) This part follows readily from part a), using \eqref{Eq:App-DualFcn} and the definition of the subgradient for $\psi(\bm\zeta)$ at $\bm\zeta(\ell)$ [cf. \eqref{Eq:App-SubgradDef-y}].

c) We have from \eqref{Eq:App-AsyncSgUpdate} for all $\bm\theta\geq \bm 0$ that
\begin{IEEEeqnarray}{l}
\left\|\bm\zeta(\ell+1) - \bm\theta \right\|^2 \negmedspace=\negmedspace \left\|\left[\bm\zeta(\ell)+\epsilon\left(\sg(\ell)+\sph(\tau(\ell))\right)\right]^+ - \bm\theta \right\|^2.
\IEEEeqnarraynumspace
\end{IEEEeqnarray}
Due to the nonexpansive property of the projection, it follows that
\begin{IEEEeqnarray}{rCl}
\left\|\bm\zeta(\ell+1) - \bm\theta \right\|^2 & \leq & \left\|\bm\zeta(\ell)+\epsilon\left(\sg(\ell)+\sph(\tau(\ell))\right) - \bm\theta \right\|^2 \nonumber\\
 & = & \left\|\bm\zeta(\ell) - \bm\theta \right\|^2 
 + \epsilon^2 \left\|\sg(\ell)+\sph(\tau(\ell)) \right\|^2 \nonumber\\
 & & +\: 2\epsilon\left(\sg(\ell)+\sph(\tau(\ell))\right)^T \left(\bm\zeta(\ell) - \bm\theta \right).
 \IEEEeqnarraynumspace
 \label{Eq:Pf-intdualiter}
\end{IEEEeqnarray}
Introducing \eqref{Eq:App-epssg-all} and \eqref{Eq:App-SubgradNormBd} into \eqref{Eq:Pf-intdualiter}, \eqref{Eq:App-basiter} follows.
\end{IEEEproof}

The main convergence results for the synchronous and asynchronous subgradient methods are given by Propositions \ref{dualconv} and \ref{dualasy}, respectively. Using Lemma \ref{Lem:DualIterRel},  Proposition \ref{dualasy} is proved next.

\begin{IEEEproof}[Proof of Proposition \ref{dualasy}]
a) Let $\bm\zeta^*$ be an arbitrary dual solution. With $\gx_i$ and $\gp_i$ denoting the $i$-th entries of $\gx$ and $\gp$, respectively, define
\begin{equation}
\delta:=\min_{1\leq i \leq d} \bigl\{-\gx_i (\av') -\exv[\gp_i(\bfp'(\bfh),\bfh)]\bigr\}
\label{Eq:Pf-slatviol}
\end{equation}
where $\av'$ and $\bfp'$ are the strictly feasible variables in AS\ref{As:Slater}.
Note that $\delta>0$ due to AS\ref{As:Slater}.

We show that the following relation holds for all $\ell \geq 1$:
\begin{multline}
\left\|\bm\zeta(\ell)-\bm\zeta^*\right\| \leq \max\Bigl\{\left\|\bm\zeta(1)-\bm\zeta^*\right\|, \\
\frac{1}{\delta}(\mathsf D - f(\av'))+\frac{\epsilon G^2}{2\delta}+\frac{2\epsilon D G \bar G}{\delta}+\|\bm\zeta^*\|+\epsilon G\Bigr\}.
\label{Eq:Pf-DualIterBd1}
\end{multline}

Eq. \eqref{Eq:Pf-DualIterBd1} implies that the sequence of Lagrange multipliers $\{\bm\zeta(\ell)\}$ is bounded, because the optimal dual set is bounded (cf. Proposition \ref{Prop:ZeroGap}). Next, \eqref{Eq:Pf-DualIterBd1} is shown by induction.

It obviously holds for $\ell=1$. Assume it holds for some $\ell\in\mathds{N}$. It is proved next that it holds for $\ell+1$. Two cases are considered, depending on the value of $\varrho(\bm\zeta(\ell))$.

\emph{Case 1:} $\varrho(\bm\zeta(\ell))>\mathsf D +\epsilon G^2/2+2\epsilon D G \bar G$. Then eq.~\eqref{Eq:App-basiter} with $\bm\theta=\bm\zeta^*$ and $\varrho(\bm\zeta^*)=\mathsf D$ becomes
\begin{multline}
\left\|\bm\zeta(\ell+1)- \bm\zeta^*\right\|^2 \leq \left\|\bm\zeta(\ell)-\bm\zeta^*\right\|^2\\ 
-\: 2\epsilon \left[\varrho(\bm\zeta(\ell)-\mathsf D - \epsilon {\bar G}^2/2 - 2 \epsilon DG\bar G \right].
\label{Eq:Pf-case2}
\end{multline}
The square-bracketed quantity in \eqref{Eq:Pf-case2} is positive due to the assumption of Case 1. Then \eqref{Eq:Pf-case2} implies that $||\bm\zeta(\ell+1)- \bm\zeta^*||^2 < ||\bm\zeta(\ell)-\bm\zeta^*||^2$, and the desired relation holds for $\ell+1$.

\emph{Case 2:} $\varrho(\bm\zeta(\ell))\leq \mathsf D +\epsilon G^2/2+2\epsilon D G \bar G$. It follows from~\eqref{Eq:App-AsyncSgUpdate}, the nonexpansive property of the projection, the triangle inequality, and the bound \eqref{Eq:App-SubgradNormBd} that 
\begin{IEEEeqnarray}{rCl}
\left\|\bm\zeta(\ell+1)- \bm\zeta^*\right\| & \leq & \left\|\bm\zeta(\ell)
+\epsilon \bigl(\sg(t)+\sph(\tau(\ell))\bigr)-\bm\zeta^*\right\| 
\IEEEyessubnumber\IEEEeqnarraynumspace\\
& \leq & \|\bm\zeta(\ell)\| + \|\bm\zeta^*\| + \epsilon G 
\IEEEyessubnumber\label{Eq:Pf-case1}
\end{IEEEeqnarray}

Next, a bound on $||\bm\zeta(\ell)||$ is developed. Specifically, it holds due to the definition of the dual function [cf. \eqref{Eq:App-DualFcn}] that 
\begin{IEEEeqnarray}{rCl}
\varrho(\bm\zeta(\ell)) & = &\max_{\av\in\boxv,\,\bfp\in\setp} 
\bigl\{f(\av) - \bm\zeta^T(\ell) \bigl(\gx(\av)+\exv[\gp(\bfp(\bfh),\bfh)]\bigr)\bigr\} 
\IEEEnonumber\\
& \geq & f(\av') - \bm\zeta^T(\ell) \bigl(\gx(\av')+\exv[\gp(\bfp'(\bfh),\bfh)]\bigr).
\IEEEeqnarraynumspace\label{Eq:Pf-dualwithslat}
\end{IEEEeqnarray}
Rewriting the inner product in \eqref{Eq:Pf-dualwithslat} using the entries of the corresponding vectors and substituting \eqref{Eq:Pf-slatviol} into \eqref{Eq:Pf-dualwithslat} using $\bm\zeta\geq\bm{0}$, it follows that
\begin{IEEEeqnarray}{rCl}
\delta \sum_{i=1}^d \bm\zeta_i(\ell) & \leq & 
-\sum_{i=1}^d \bm\zeta_i^T(\ell) \bigl(\gx_i(\av')+\exv[\gp_i(\bfp'(\bfh),\bfh)]\bigr) 
\IEEEnonumber\\
& \leq & \varrho(\bm\zeta(\ell)) - f(\av').
\label{Eq:Pf-norm1bd}
\end{IEEEeqnarray}
Using $\|\bm\zeta(\ell)\| \leq \sum_{i=1}^d \bm\zeta_i(\ell)$ into \eqref{Eq:Pf-norm1bd}, the following bound is obtained:
\begin{equation}
\|\bm\zeta(\ell)\| \leq \frac{1}{\delta} (\varrho(\bm\zeta(\ell)) - f(\av')).
\label{Eq:Pf-norm2bd}
\end{equation}

Introducing \eqref{Eq:Pf-norm2bd} into \eqref{Eq:Pf-case1} and using the assumption of Case 2, the desired relation \eqref{Eq:Pf-DualIterBd1} holds for $\ell+1$.


b) Set $\bm\theta=\bm\zeta^*$ and $\varrho(\bm\theta)=\varrho(\bm\zeta^*)=\mathsf D$ in \eqref{Eq:App-basiter}:
\begin{IEEEeqnarray}{rCl}
\left\|\bm\zeta(\ell+1)-\bm\zeta^*\right\|^2 & \leq & \left\|\bm\zeta(\ell)-\bm\zeta^*\right\|^2 
+ \epsilon^2 {G}^2+ 4 \epsilon^2 DG\bar G \nonumber\\
 & & + \: 2\epsilon \left[\mathsf D - \varrho(\bm\zeta(\ell))\right]. 
 \IEEEeqnarraynumspace\label{Eq:Pf-basiteropt-1}
\end{IEEEeqnarray}
Summing the latter for $\ell=1,\ldots,s$, and introducing the quantity $\min_{1\leq \ell \leq s} \varrho(\bm\zeta(\ell))$, it follows that
\begin{IEEEeqnarray}{rCl}
\left\|\bm\zeta(\ell+1)-\bm\zeta^*\right\|^2 
 & \leq & \left\|\bm\zeta(1)-\bm\zeta^*\right\|^2  + s\epsilon^2 {G}^2+ 4 s\epsilon^2 DG\bar G \nonumber\\
 & & +\: 2s\epsilon \mathsf D - 2s\epsilon \min_{1\leq \ell \leq s} \varrho(\bm\zeta(\ell)).
  \IEEEeqnarraynumspace\label{Eq:Pf-basiteropt-2} 
\end{IEEEeqnarray}
%
Substituting the left-hand side of~\eqref{Eq:Pf-basiteropt-2} with 0, rearranging the resulting inequality, and dividing by $2\epsilon s$, we obtain
\begin{IEEEeqnarray}{rCl}
 \min_{1\leq \ell \leq s} \varrho(\bm\zeta(\ell)) \negmedspace & \leq & \negmedspace\mathsf D 
+\frac{\epsilon {G}^2}{2} + 2 \epsilon DG\bar G +\frac{\left\|\bm\zeta(1)-\bm\zeta^*\right\|^2}{2\epsilon s}.
\IEEEeqnarraynumspace
  \label{Eq:Pf-basiteropt-3} 
\end{IEEEeqnarray}

Now, note that $\lim_{s\rightarrow\infty}\min_{1\leq \ell \leq s} \varrho(\bm\zeta(\ell))$ exists, because $\min_{1\leq \ell \leq s} \varrho(\bm\zeta(\ell))$ is monotone decreasing in $s$ and lower-bounded by $\mathsf D$, which is finite. Moreover, $\lim_{s\rightarrow\infty}{\left\|\bm\zeta(1)-\bm\zeta^*\right\|^2}/({2\epsilon s})=0$, because $\bm\zeta^*$ is bounded. Thus, taking the limit as $s\rightarrow\infty$ in \eqref{Eq:Pf-basiteropt-3}, yields \eqref{asyncbestdualval}.
\end{IEEEproof}

Note that the sequence of Lagrange multipliers in the synchronous algorithm \eqref{Eq:App-SyncSgUpdate} is bounded. This was shown for convex primal problems in \cite[Lemma 3]{nedic07}. Interestingly, the proof also applies in the present case since AS\ref{As:box}-AS\ref{As:Slater} hold and imply finite optimal $\mathsf{P} = \mathsf{D}$. 
Furthermore, Proposition~\ref{dualconv} for the synchronous method follows from \cite[Prop.~8.2.3]{Bertsekas-ConvexAnalysis}, \cite{ale09}. 

Next, the convergence of primal variables through running averages is considered. The following lemma collects the intermediate results for the averaged sequence $\{\bar\av(s)\}$ [cf. \eqref{runav}], and is used to establish convergence for the generic problem \eqref{Eq:App-OptPr-All} with asynchronous subgradient updates as in \eqref{Eq:App-AsyncSgUpdate}. Note that $\bar\av(s)\in\boxv$, $s \geq 1$, because \eqref{runav} represents a convex combination of the points $\{\av(1),\ldots,\av(s)\}$.
\begin{lem}
\label{Lem:AvgSeqPropert}
Under AS\ref{As:box}-AS\ref{As:BddDelay} with $\bm\zeta^*$ denoting an optimal Lagrange multiplier vector, there exists a sequence $\{\mathring\bfp(.;s)\}$ in $\setp$ such that for any $s\in\mathds{N}$, it holds that
\begin{subequations}
\begin{align}
& \hspace{-0.16cm}\textnormal{a)~} \left\|\left[\gx(\bar\av(s))+\exv\left[\gp(\mathring\bfp(\bfh;s),\bfh)\right]\right]^+  \right\| \leq \frac{\left\|\bm\zeta(s+1)\right\|}{\epsilon s} \label{Eq:App-ConstrViolBd}\\
& \hspace{-0.16cm}\textnormal{b)~} f(\bar\av(s)) \geq \mathsf{D} -\dfrac{\left\|\bm\zeta(1)\right\|^2}{2\epsilon s} - \dfrac{\epsilon {G}^2}{2} - 2 \epsilon D G \bar G \label{Eq:App-CostLowBd-sgnorm}\\ 
& \hspace{-0.16cm}\textnormal{c)~} f(\bar\av(s)) \leq \mathsf{D} +\left\|{\bm\zeta^*}\right\|
 \left\|\left[ \gx(\bar\av(s))+\exv[\gp(\mathring\bfp(\bfh;s),\bfh)] \right]^+ \right\|.
\label{Eq:App-CostUppBd}
\end{align}
\end{subequations}
\end{lem}

Eq. \eqref{Eq:App-ConstrViolBd} is an upper bound on the constraint violation, while \eqref{Eq:App-CostLowBd-sgnorm} and \eqref{Eq:App-CostUppBd} provide lower and upper bounds on the objective function at $\bar\av(s)$. Lemma \ref{Lem:AvgSeqPropert} relies on Lemma \ref{Lem:ConvexRange} and the fact that the averaged sequence $\{\bar\av(s)\}$ is generated from maximizers of the Lagrangian $\{\av(\ell)\}$ that are \emph{not} outdated.

\begin{IEEEproof}[Proof of Lemma \ref{Lem:AvgSeqPropert}]
a) It follows from \eqref{Eq:App-AsyncSgUpdate} that
\begin{equation}\label{zetalpo}
\bm\zeta(\ell+1) \geq \bm\zeta(\ell)+\epsilon\left(\sg(\ell)+\sph(\tau(\ell))\right).
\end{equation}
Summing \eqref{zetalpo} over $\ell=1,\ldots,s$, using $\bm\zeta(1)\geq \bm 0$, and dividing by $2\epsilon s$, it follows that
\begin{IEEEeqnarray}{rCl}
\frac{1}{s}\sum_{\ell=1}^{s}\sg(\ell)+\frac{1}{s} \sum_{\ell=1}^{s} \sph(\tau(\ell)) & \leq & \frac{\bm\zeta(s+1)}{\epsilon s}.
\label{Pf-IneqAvgSg}
\end{IEEEeqnarray}

Now, recall the definitions of the subgradients $\sg(\ell)$ and $\sph(\tau(\ell))$ in \eqref{Eq:App-SubgradDef-ALL}. Due to the convexity of $\gx(\cdot)$, it holds that
\begin{equation}
\label{Pf-IneqAvgy}
\gx(\bar\av(s))\leq \frac{1}{s}\sum_{\ell=1}^{s} \gx(\av(\ell))=\frac{1}{s}\sum_{\ell=1}^{s} \sg(\ell).
\end{equation}
Due to Lemma~\ref{Lem:ConvexRange}, there exists $\mathring\bfp(\bfh;s)$ in $\setp$ such that
\begin{equation}
\label{Pf-IneqAvgsph}
\exv[\gp(\mathring\bfp(\bfh;s),\bfh)] \negthinspace= \negthinspace \frac{1}{s}\sum_{\ell=1}^{s} \exv[\gp(\bfp(\bfh;\tau(\ell)),\bfh)] \negthinspace = \negthinspace \frac{1}{s} \sum_{\ell=1}^{s} \sph(\tau(\ell)).
\end{equation}

Combining \eqref{Pf-IneqAvgSg}, \eqref{Pf-IneqAvgy}, and \eqref{Pf-IneqAvgsph}, it follows that 
\begin{equation}\label{gybars}
\gx(\bar\av(s)) + \exv[\gp(\mathring\bfp(\bfh;s),\bfh)] \leq \frac{\bm\zeta(s+1)}{\epsilon s}.
\end{equation}
Using $\bm\zeta(s+1)\geq \bm 0$ and the fact that $[.]^+$ is a nonnegative vector, \eqref{Eq:App-ConstrViolBd} follows easily from~\eqref{gybars}.

b) Due to the concavity of $f(\cdot)$, it holds that $f(\bar\av(s)) \geq \frac{1}{s} \sum_{\ell=1}^{s} f(\av(\ell))$. 
Adding and subtracting the same terms, $\bm\zeta^T(\ell)\gx(\av(\ell))$ and $\bm\zeta^T(\ell)\exv[\gp(\bfp(\bfh;\tau(\ell)),\bfh)]$ for $\ell=1,\ldots,s$, to the right-hand side of the latter, and using $f(\av(\ell))-\bm\zeta^T(\ell)\gx(\av(\ell))=\psi(\bm\zeta(\ell))$ [cf.~\eqref{Eq:App-MaxSeq-v} and~\eqref{Eq:App-DualFcn}], it follows that
%
%
\begin{multline}
f(\bar\av(s)) \geq 
\frac{1}{s} \sum_{\ell=1}^{s} \Bigl[\psi(\bm\zeta(\ell)) - \bm\zeta^T(\ell)\exv[\gp(\bfp(\bfh;\tau(\ell)),\bfh)] \Bigr]\\
+\: \frac{1}{s} \sum_{\ell=1}^{s} \bm\zeta^T(\ell)\bigl(\gx(\av(\ell))+\exv[\gp(\bfp(\bfh;\tau(\ell)),\bfh)] \bigr).
\label{Eq:Pf-CostLowBd-2InProd}
\end{multline}

Now recall that $\exv[\gp(\bfp(\bfh;\tau(\ell)),\bfh)]=\sph(\tau(\ell))$ [cf. \eqref{Eq:App-SubgradDef-ph}]. Thus, it holds that
\begin{multline}
-\zeta^T(\ell)\exv[\gp(\bfp(\bfh;\tau(\ell)),\bfh)] \\
=-\bm\zeta^T(\tau(\ell))\sph(\tau(\ell))+\sph^T(\tau(\ell))\bigl[\bm\zeta(\tau(\ell))-\bm\zeta(\ell)\bigr].
\label{Eq:Pf-2ndInProdExpand}
\end{multline}
The first term in the right-hand side of \eqref{Eq:Pf-2ndInProdExpand} is $\phi\bigl(\bm\zeta(\tau(\ell))\bigr)$ ([cf. \eqref{Eq:App-MaxSeq-ph} and \eqref{Eq:App-DualFcn}]. The second term can be lower-bounded using Lemma \ref{Lem:DualIterRel}(a) with $\bm\theta=\bm\zeta(\tau(\ell))$. Then, \eqref{Eq:Pf-2ndInProdExpand} becomes
\begin{equation}
-\zeta^T(\ell)\exv[\gp(\bfp(\bfh;\tau(\ell)),\bfh)] \geq \phi(\bm\zeta(\ell)) - 2\epsilon D G \bar G
\label{Eq:Pf-2ndInProdwithDual}
\end{equation}
Using \eqref{Eq:Pf-2ndInProdwithDual} into \eqref{Eq:Pf-CostLowBd-2InProd} and $\psi(\bm\zeta(\ell))+\phi(\bm\zeta(\ell))=\varrho(\bm\zeta(\ell))\geq \mathsf D$, it follows that 
\begin{multline}
f(\bar\av(s))  \geq \mathsf D  - 2\epsilon D G \bar G \\
 +\: \frac{1}{s} \sum_{\ell=1}^{s} \bm\zeta^T(\ell)\bigl(\gx(\av(\ell))+\exv[\gp(\bfp(\bfh;\tau(\ell)),\bfh)] \bigr).
\label{Eq:Pf-CostLowBd-1InProd}
\end{multline}

Moreover, it follows from~\eqref{Eq:App-AsyncSgUpdate} and the nonexpansive property of the projection that 
\begin{IEEEeqnarray*}{rCl}
\|\bm\zeta(\ell+1)\|^2 & \leq & \|\bm\zeta(\ell)\|^2 \\
 & & +\: 2\epsilon\bm\zeta^T(\ell) \left( \gx(\av(\ell))+\exv[\gp(\bfp(\bfh;\tau(\ell)),\bfh)] \right) \\
 & & +\: \epsilon^2 \left\| \gx(\av(\ell))+\exv[\gp(\bfp(\bfh;\ell),\bfh)] \right\|^2. 
 \IEEEeqnarraynumspace\IEEEyesnumber\label{Eq:Pf-SqExpand}
\end{IEEEeqnarray*}
Summing \eqref{Eq:Pf-SqExpand} for $\ell=1,\ldots,s$, dividing by $2\epsilon s$, and introducing the bound \eqref{Eq:App-SubgradNormBd} on the subgradient norm yield
\begin{multline}
\frac{1}{s}\sum_{\ell=1}^{s} \bm\zeta^T(\ell) \bigl(\gx(\av(\ell))+\exv[\gp(\bfp(\bfh;\ell),\bfh)] \bigr) \\
 \geq -\frac{\epsilon G^2}{2}+ \frac{\|\bm\zeta(s+1)\|^2-\|\bm\zeta(1)\|^2}{2\epsilon s}.
 \label{Eq:Pf-3rdInProdBd}
\end{multline}
Using \eqref{Eq:Pf-3rdInProdBd} into~\eqref{Eq:Pf-CostLowBd-1InProd} together with $\|\bm\zeta(s+1)\|^2\geq 0$, one arrives readily at \eqref{Eq:App-CostLowBd-sgnorm}.

c) Let $\bm\zeta^*$ be an optimal dual solution.
It holds that
\begin{IEEEeqnarray}{rCl}
f(\bar\av(s)) & = & f(\bar\av(s))
    -  {\bm\zeta^*}^T\bigl(\gx(\bar\av(s))+\exv[\gp(\mathring\bfp(\bfh;s),\bfh)] \bigr) \notag\\
  & &  +\: {\bm\zeta^*}^T\bigl(\gx(\bar\av(s))+\exv[\gp(\mathring\bfp(\bfh;s),\bfh] \bigr)
   \label{Eq:Pf-fxbarEqualiii}
\end{IEEEeqnarray}
where $\mathring\bfp(\bfh;s)$ was defined in part (a) [cf. \eqref{Pf-IneqAvgsph}].

By the definitions of $\mathsf D$ and $\bm\zeta^*$ [cf. \eqref{Eq:App-DualPr}], and the dual function [cf. \eqref{Eq:App-DualFcn}], it holds that 
\begin{equation}
\mathsf D=\varrho(\bm\zeta^*)=\max_{\av\in\boxv,\bfp\in\setp} \mathcal L(\av, \bfp, \bm\zeta^*)\geq \mathcal L(\bar\av, \mathring\bfp, \bm\zeta^*). 
\end{equation}
Substituting the latter into \eqref{Eq:Pf-fxbarEqualiii}, it follows that
\begin{equation}
f(\bar\av(s))  \leq  \mathsf D
    +{\bm\zeta^*}^T\bigl(\gx(\bar\av(s))+\exv[\gp(\mathring\bfp(\bfh;s),\bfh)] \bigr).
\label{Eq:Pf-CostUpBd-WithInProd}
\end{equation}
Because $\bm\zeta^*\geq \bm{0}$ and $\bm\theta \leq [\bm\theta]^+ $ for all $\bm\theta$, \eqref{Eq:Pf-CostUpBd-WithInProd} implies that 
\begin{equation}
f(\bar\av(s))  \leq  \mathsf D
    +{\bm\zeta^*}^T\bigl[ \gx(\bar\av(s))+\exv[\gp(\mathring\bfp(\bfh;s),\bfh)] \bigr]^+.
\end{equation}
Applying the Cauchy-Schwartz inequality to the latter, \eqref{Eq:App-CostUppBd} follows readily.
\end{IEEEproof}

Using Lemma \ref{Lem:AvgSeqPropert}, the main convergence results for the synchronous and asynchronous subgradient methods are given correspondingly by Propositions \ref{primconv} and \ref{primalasy}, after substituting
\begin{equation}
\mathbf{q}(\bar{\mathbf{y}}(s),\mathring{\mathbf{p}}(\mathbf{h};s)) = \gx(\bar\av(s))+\exv[\gp(\mathring\bfp(\bfh;s)].
\end{equation}

\begin{IEEEproof}[Proof of Proposition \ref{primalasy}]
a) Take limits on both sides of~\eqref{Eq:App-ConstrViolBd} as $s\rightarrow\infty$, and use the boundedness of $\{\bm\zeta(s)\}$.

b) Using $\mathsf{P} = \mathsf{D}$ and taking the $\liminf$ in~\eqref{Eq:App-CostLowBd-sgnorm}, we obtain~\eqref{inf-primaconst}. Moreover, using $\mathsf{P} = \mathsf{D}$, \eqref{Eq:App-ConstrViolBd}, the boundedness of $\|\bm\zeta^*\|$, and taking $\limsup$ in \eqref{Eq:App-CostUppBd}, \eqref{sup-primaconst} follows.
\end{IEEEproof}

%

\bibliographystyle{IEEEtran}
\bibliography{IEEEabrv,separation-biblio}
\vspace{-1cm}
\begin{biography}[]{Ketan Rajawat}
(S'06) received his B.Tech.and M.Tech.degrees in Electrical Engineering from Indian Institute of Technology Kanpur in 2007. Since August 2007, he has been working towards his Ph.D. degree at the Department of Electrical and Computer Engineering, University of Minnesota, Minneapolis. His current research focuses on network coding and optimization in wireless networks.
\end{biography}
\vspace{-1cm}
\begin{biography}[]{Nikolaos Gatsis}
received the Diploma degree in
electrical and computer engineering from the University
of Patras, Patras, Greece in 2005 with honors.
Since September 2005, he has been working
toward the Ph.D. degree with the Department of
Electrical and Computer Engineering, University of
Minnesota, Minneapolis, MN. His research interests
include cross-layer designs, resource allocation, and
signal processing for wireless networks.
\end{biography}
\vspace{-1cm}
\begin{biography}[]{G. B.  Giannakis}
(F'97) received his Diploma in Electrical 
Engr. from the Ntl. Tech. Univ. of Athens, Greece, 1981. From 
1982 to 1986 he was with the Univ. of Southern California (USC), 
where he received his MSc. in Electrical Engineering, 1983, MSc. 
in Mathematics, 1986, and Ph.D. in Electrical Engr., 1986. Since 
1999 he has been a professor with the Univ. of Minnesota, where 
he now holds an ADC Chair in Wireless Telecommunications in the 
ECE Department, and serves as director of the Digital Technology
Center. 

His general interests span the areas of communications, networking 
and statistical signal processing - subjects on which he has 
published more than 300 journal papers, 500 conference papers, 
two edited books and two research monographs. Current research 
focuses on compressive sensing, cognitive radios, network coding, 
cross-layer designs, mobile ad hoc networks, wireless sensor and 
social networks. He is the (co-) inventor of 18 to 20 patents issued, and 
the (co-) recipient of seven paper awards from the IEEE Signal 
Processing (SP) and Communications Societies, including the G. Marconi 
Prize Paper Award in Wireless Communications. He also received 
Technical Achievement Awards from the SP Society (2000), from 
EURASIP (2005), a Young Faculty Teaching Award, and the G. W. Taylor 
Award for Distinguished Research from the University of Minnesota. 
He is a Fellow of EURASIP, and has served the IEEE in a number of posts, 
including that of a Distinguished Lecturer for the IEEE-SP Society. 

\end{biography}

\end{document}